\documentclass{jpp}

\usepackage{graphicx,bm,color}
\usepackage{natbib}

\newcommand{\EQ}{\begin{equation}}
\newcommand{\EN}{\end{equation}}
\newcommand{\EQA}{\begin{eqnarray}}
\newcommand{\ENA}{\end{eqnarray}}

\newcommand{\EEq}[1]{Equation~(\ref{#1})}
\newcommand{\Eq}[1]{Eq.~(\ref{#1})}
\newcommand{\Eqs}[2]{Eqs.~(\ref{#1}) and~(\ref{#2})}

\newcommand{\Eqss}[2]{Eqs.~(\ref{#1})--(\ref{#2})}
\newcommand{\EEqss}[2]{Equations~(\ref{#1})--(\ref{#2})}

\newcommand{\App}[1]{Appendix~\ref{#1}}

\newcommand{\Fig}[1]{Fig.~\ref{#1}}

{}
{}
{}

{}
{}
{}
{}
{}
{}
{}
{}
{}
{}
{}
{}
{}
{}
{}
{}
{}

{}
{}
{}

{}

{}
{}

\def\talp{\tau_{\alpha}}
\newcommand{\beq}{\begin{equation}}
\newcommand{\eeq}{\end{equation}}

\newcommand{\bfb}{\mbox{\boldmath $b$}}

\newcommand{\bfh}{\mbox{\boldmath $h$}}
\newcommand{\bfk}{\mbox{\boldmath $k$}}

\newcommand{\bfv}{\mbox{\boldmath $v$}}
\newcommand{\bfx}{\mbox{\boldmath $x$}}

\newcommand{\bfr}{\mbox{\boldmath $r$}}

\newcommand{\bfB}{\mbox{\boldmath $B$}}

\newcommand{\bfE}{\mbox{\boldmath $E$}}

\newcommand{\bfH}{\mbox{\boldmath $H$}}
\newcommand{\bfK}{\mbox{\boldmath $K$}}
\newcommand{\bfM}{\mbox{\boldmath $M$}}
\newcommand{\bfR}{\mbox{\boldmath $R$}}
\newcommand{\bfU}{\mbox{\boldmath $U$}}
\newcommand{\bfV}{\mbox{\boldmath $V$}}

\newcommand{\bfX}{\mbox{\boldmath $X$}}

\newcommand{\bfxi}{\mbox{\boldmath $\xi$}}
\newcommand{\ex}{\mbox{{\boldmath $e$}}_{1}}
\newcommand{\ey}{\mbox{{\boldmath $e$}}_{2}}
\newcommand{\ez}{\mbox{{\boldmath $e$}}_{3}}

\newcommand{\bfemf}{\mbox{\boldmath ${\cal E}$}}
\newcommand{\bnabla}{\mbox{\boldmath $\nabla$}}
\newcommand{\cross}{\mbox{\boldmath $\times$}}
\newcommand{\cendot}{\mbox{\boldmath $\cdot\,$}}
\newcommand{\rem}{{\rm Rm}}

\title[Mean field dynamo due to stochastic $\alpha$ in shear flows]{Generation of large-scale magnetic fields due to fluctuating $\alpha$ in shearing systems}

\author[Jingade, Singh \& Sridhar]%
{N\ls A\ls V\ls E\ls E\ls N\ns J\ls I\ls N\ls G\ls A\ls D\ls E$^{1,2}$
  \thanks{Email address for correspondence: naveen@rri.res.in},
N\ls I\ls S\ls H\ls A\ls N\ls T\ns K.\ns S\ls I\ls N\ls G\ls H$^{3}$
\and
S.\ns S\ls R\ls I\ls D\ls H\ls A\ls R$^{2}$
}

\affiliation{
$^{1}$Indian Institute of Science, Bangalore 560 012, India\\
$^{2}$Raman Research Institute, Sadashivanagar, Bangalore 560 080, India\\
$^{3}$Max Planck Institute for Solar System Research,
Justus-von-Liebig-Weg 3, D-37077 G\"{o}ttingen, Germany
}

\begin{document}

\maketitle

\begin{abstract}
We explore the growth of large--scale magnetic fields in a shear flow,
due to helicity fluctuations with a finite correlation time, through a study
of the Kraichnan--Moffatt model of zero--mean stochastic fluctuations 
of the $\alpha$ parameter of dynamo theory. We derive a linear integro--differential equation for the evolution of large--scale magnetic field, using the first--order smoothing approximation and the Galilean
invariance of the $\alpha$--statistics. This enables construction of
a model that is non--perturbative in the shearing rate $S$ and 
the $\alpha$--correlation time $\talp$.  After a brief review of the 
salient features of the exactly solvable white--noise limit, we consider 
the case of small but non--zero $\talp$. When the large--scale magnetic 
field varies slowly, the evolution is governed by a partial differential equation. We present modal solutions and conditions for the exponential growth rate of the large--scale magnetic field, whose drivers are
the Kraichnan diffusivity, Moffatt drift, Shear and a non--zero correlation 
time. Of particular interest is dynamo action when the $\alpha$--fluctuations are weak; i.e. when the Kraichnan diffusivity is positive.
We show that in the absence of Moffatt drift, shear does not give 
rise to growing solutions. But shear and Moffatt drift acting together can drive large scale dynamo action with growth rate $\gamma \propto |S|$.
\end{abstract}

\section{Introduction}

Magnetic fields are observed over a wide range of scales in various
astrophysical objects \citep[see, e.g.,][]{JL17}. Their origins could be the result of turbulent 
dynamo processes which can lead to field generation on scales that are
larger as well as smaller than the outer scale of underlying turbulence
\citep[see, e.g.,][]{moffatt1978,parker1979,krause1980,zeldovich1983,ruzmaikin1988,BS05}.
Of particular interest here is the subject of large-scale dynamo (LSD) which
may be studied in the framework of mean-field theory
\citep{SKR66,moffatt1978,krause1980}. The standard paradigm for LSD involves an $\alpha$-effect which arises when the background turbulence possesses mean kinetic helicity, thus
breaking the mirror symmetry of turbulence \cite[see, e.g.,][]{BS05}.
The problem becomes more interesting and complicated when the usual
$\alpha$-effect is either absent or subcritical for dynamo growth.
Mean velocity shear appears to play a vital role for LSD in such regimes
of zero/subcritical $\alpha$. As most astrophysical bodies also possess
mean differential rotation, it is natural to ask if large-scale magnetic
fields could grow in the presence of a background shear flow when $\alpha$ is a purely fluctuating quantity. 

Early ideas of stochastically varying $\alpha$ with zero mean
suggested that it causes a decrement in turbulent diffusion
\citep{Kra76,moffatt1978}. A number of subsequent studies then considered
fluctuating $\alpha$ as an important ingredient for the evolution
of magnetic fields in objects, such as, the Sun \citep{Sil00,Pro07},
the accretion disks \citep{VB97}, galaxies \citep{Sok97,SurSub09}.
Numerical demonstration of the shear dynamo problem
\citep{You08a,You08b,BRRK08,SJ15} where large--scale magnetic fields were
generated due to non-helically forced turbulence in shear flows,
and failure to understand these in terms of simple ideas involving
shear--current effect \citep{KR08,RK08,SS09a,SS09b,SS10,SS11,KSS12},
brought the focus on stochastic $\alpha$
which could potentially lead to the dynamo action generically in
shearing systems \citep{HMS11,McW12,MB12,Pro12,RP12,SS14}.
There is still a need to verify the model predictions for the growth
of first moment of the mean magnetic field in such systems by performing
more simulations.

\cite{SB15a,SB15b} recently proposed a new mechanism, called the
magnetic shear current effect, which leads to the generation of a large scale magnetic field due to the combined action of shear and small scale
magnetic fluctuations, if these are sufficiently strong and are near
equipartition levels of turbulent motions. Such strong magnetic fluctuations are expected to be naturally present due to small scale dynamo (SSD) action in astrophysical plasmas, which typically have large magnetic Reynolds number ($\rem$). This new effect thus raises the interesting possibility of the excitation of LSD due to SSD in presence of shear, and it challenges an understanding where SSD in high-$\rem$ systems is thought to weaken the LSD, which could survive only when SSD is suppressed due to shear
\citep{TC13,Pon16,Nig17}; but see also \cite{Kol11,SRB17} where
it is found that the shear supports and even enhances the growth rate of SSD. However, we are here more concerned with the excitation of a large--scale shear dynamo, quite independent of any small--scale dynamo or
strong magnetic fluctuations, which are both absent
in most numerical simulations that are relevant. These simulations
typically had $\rem$ which were subcritical for SSD and the only source
of magnetic fluctuations was due to the tangling of large--scale magnetic
fields \citep{rogachevskii2007}, and therefore these fluctuations
could never be too strong in the kinematic regime of LSD.

In the present paper we explore the possibility of large-scale dynamo
action in presence of background shear flow, due an $\alpha$ that varies
stochastically in space and time, with vanishing mean.
Here we generalize the earlier work
by \cite{SS14}, hereafter SS14, by including the full resistive term in
determining the turbulent electromotive force (EMF). Such an extension
in the absence of shear was done in \cite{S16}.
In Section~2 we define our model by
writing dynamo equations in shearing coordinates. Integro--differential
equation governing the evolution of the large-scale magnetic field
is derived under FOSA in Section~3. This is non-perturbative in shearing
rate $S$ and the correlation time $\talp$. Here we briefly review the
exactly solvable limit of white--noise $\alpha$ fluctuations.
In Section~4 we reduce the evolution equation into a partial differential
equation (PDE) for axisymmetric mean magnetic fields, by assuming small
but non-zero $\talp$. Dispersion relation giving the growth rate is then
determined in Section~5 where we present our results in different
parameter regimes. We then discuss our findings and conclude in Section~6.

\section{The model}

\noindent
Let us begin with the standard dynamo equation in the presence of
a background linear shear flow, $\bfV=SX_1\ey$,where meso-scale magnetic field $\bfB$ evolves according to \citep[see,][]
{moffatt1978,krause1980,BS05,SS14}:
\beq
\left(\frac{\partial}{\partial\tau} \,+\,
SX_1\frac{\partial}{\partial X_2}\right)\!\bfB\;-\;
SB_1\ey \;=\; \bnabla\cross\left[\,\alpha(\bfX, \tau) \bfB\,\right]\,+\,
\eta_T\bnabla^2\bfB\,;
\quad \bnabla\cendot\bfB \;=\; 0\,. 
\label{mesoB}
\eeq
Here we follow the same notation as in \cite{SS14} where the position
vector is denoted by $\bfX = (X_1, X_2, X_3)$ with components given
in a fixed orthonormal frame $(\ex, \ey, \ez)$, and $\tau$ is the
time variable. The shear rate, $S$, and total diffusivity,
$\eta_T$, are treated as constant parameters, whereas $\alpha(\bfX, \tau)$ providesa measure of meso-scale kinetic helicity of turbulence.
We recall that \Eq{mesoB} governing the dynamics of meso-scale
magnetic field is obtained by averaging over an ensemble of random
velocity fields, $\{\bfv(\bfX, \tau)\}$, which are assumed to have
zero-mean isotropic fluctuations, uniform and constant kinetic energy
density per unit mass, and slow helicity fluctuations.

We employ here the double-averaging scheme \citep{Kra76,Mof83,Sok97}
under which $\alpha(\bfX, \tau)$ itself is a random variable
of space and time, thus making \Eq{mesoB} a stochastic partial
differential equation. It is drawn from a superensemble with zero
mean, $\overline{\alpha(\bfX, \tau)}=0\,$. It's statistical properties
are given below in \Eq{aacorr}.
Next, we separate the meso-scale field, $\bfB=\overline{\bfB}+\bfb$,
into large-scale, $\overline{\bfB}$, and fluctuating, $\bfb$,
components, where the superensemble average of $\bfb$ vanishes,
i.e., $\overline{\bfb}={\bf 0}$. Governing equation for the large-scale
magnetic field $\overline{\bfB}$ can thus be obtained by Reynolds
averaging the \Eq{mesoB} over the superensemble:
\begin{eqnarray}
\left(\frac{\partial}{\partial\tau} \,+\, SX_1\frac{\partial}{\partial X_2}\right)\overline{\bfB} \;-\; S\overline{B_1}\ey 
&\;=\;& \bnabla\cross\overline{\bfemf} \;+\; \eta_T\bnabla^2\overline{\bfB}\,,\qquad\qquad \bnabla\cendot\overline{\bfB} \;=\; 0\,,
\label{lsb}\\[2ex]
\mbox{where}\qquad\overline{\bfemf} &\;=\;& \overline{\alpha(\bfX, \tau)\bfb(\bfX, \tau)}\,.
\label{memf}
\end{eqnarray}
In order to determine the mean electromotive force (EMF), we must solve
for the fluctuating field $\bfb$, which evolves as:
\begin{eqnarray}
\left(\frac{\partial}{\partial\tau} \,+\, SX_1\frac{\partial}{\partial X_2}\right)\bfb \;-\; Sb_1\ey &\;=\;& \bnabla\cross\left[\,\alpha \overline{\bfB}\,\right] \,+\,\bnabla\cross\left[\,\alpha\bfb-\overline{\alpha\bfb}\right]\,+\,\eta_T\bnabla^2\bfb\,, \nonumber\\[2ex]
\bnabla\cendot\bfb \;=\; 0\,,&& \mbox{with initial condition}\quad\bfb(\bfX, 0)=\bf0\,.
\label{fb}
\end{eqnarray}
As \Eqs{lsb}{fb} involve inhomogeneous terms, it is convenient to solve
these in shearing frame where shearing coordinates $(\bfx, t)$
are expressed in terms of the lab coordinates $(\bfX, \tau)$ as
\citep[see,][]{SS09a,SS10}:
\beq
x_1\,=\,X_1\,;\qquad x_2\,=\,X_2-S\tau X_1\,;\qquad x_3\,=\,X_3\,;\qquad t\,=\,\tau\,.
\label{tr-sh}
\eeq
The inverse transformation is:
\beq
X_1\,=\,x_1\,;\qquad X_2\,=\,x_2+St x_1\,;\qquad X_3\,=\,x_3\,;\qquad \tau\,=\,t\,.
\label{invtr-sh}
\eeq
Now we can write \Eqss{lsb}{fb} in terms of new fields that are functions of
$\bfx$ and $t$: $\overline{\bfH}(\bfx, t)=\overline{\bfB}(\bfX, \tau)$;
$\bfh(\bfx, t)=\bfb(\bfX, \tau)$; $a(\bfx, t)=\alpha(\bfX, \tau)$;
and $\overline{\bfE}(\bfx, t)=\overline{\bfemf}(\bfX, \tau)\,$.
\EEqss{lsb}{fb} then take the form \citep{SS14}:
\beq
\frac{\partial \overline{\bfH}}{\partial t} \;-\; S\overline{H_1}\ey \;=\;
\bnabla\cross\overline{\bfE} \,+\, \eta_T\bnabla^2\overline{\bfH}
\,,\qquad\bnabla\cendot\overline{\bfH} \;=\; 0\,,
\qquad\overline{\bfE} \;=\; \overline{a\bfh}\,;
\label{lsbsh}
\eeq

\begin{eqnarray}
\frac{\partial \bfh}{\partial t} \;-\; Sh_1\ey &\;=\;&
\bnabla\cross\left[\,a\overline{\bfH}\,\right] \,+\,
\bnabla\cross\left[\,a\bfh-\overline{a\bfh}\right]\,+\,\eta_T\bnabla^2\bfh
\,,\nonumber\\[1ex]
\bnabla\cendot\bfh &\;=\;& 0\,,\quad \mbox{with initial condition}\quad\bfh(\bfx, 0)=\bf0\,;
\label{fbsh}\\[2ex]
\mbox{where} \qquad\bnabla &\;=\;& \frac{\partial}{\partial \bfx} \,-\, 
\ex St\frac{\partial}{\partial x_2}\qquad \mbox{is a time-dependent operator.}
\label{nabsh} 
\end{eqnarray}

We complete defining our model by specifying the statistics of $\alpha$
fluctuations. We follow the exact same approach as given in detail in
\cite{SS14} and recall here only some key relevant points:
\begin{itemize}
\item Shear flows possess a natural symmetry known as Galilean invariance,
relating the measurements of correlation functions made by comoving
observers whose origins with resepct to the lab frame translate with
the same speed as that of the linear shear flow \citep{SS09a,SS09b}.
\item Here we are more interested in time--stationary Galilean--invariant
$\alpha$ statistics, which can be expressed in the shearing
frame as (see \cite{SS14} for a derivation):
\begin{eqnarray}
\overline{a(\bfx, t) a(\bfx', t')} \;&=&\;
2{\cal A}\!\left(\bfx-\bfx'+St'(x_1-x_1')\ey\right)\,{\cal D}(t-t')\,,
\qquad\mbox{with}\\[1em]
2\int_0^{\infty}\,{\cal D}(t){\rm d}t &\;=\;& 1\,,\qquad\qquad
{\cal A}({\bf 0}) \;=\; \eta_{\alpha} \,\geq\, 0\,.
\label{aacorr}
\end{eqnarray}
The correlation time for the $\alpha$ fluctuations is defined as,
\beq
\tau_{\alpha} \;=\; 2\int_0^{\infty} \mathrm{d}t\;t\; {\cal D}(t)\,.
\label{corr-time}
\eeq
The intrinsic anisotropy of the $\alpha$ fluctuations is measured 
by the Moffatt drift velocity,
\beq
\bfV_{\!\!M} \;=\; -\left(\frac{\partial {\cal A}(\bfxi)}{\partial\bfxi}\right)_{\bfxi =\bf0} \;=\; 
\int_0^\infty
\overline{\alpha(\bfX, \tau)\bnabla\!\alpha(\bfX, 0)}\,{\rm d}\tau\,. 
\label{kmcondef}
\eeq
\end{itemize}
In the above, we noted two properties of the spatial correlation function, ${\cal A}$, namely its value $\eta_\alpha$ and gradient $\bfV_{\!\!M}$ at zero separation. But we can associate one or more length scales relating to its variation in $\bfxi$-space. In the estimates made below we use a single scale $\ell$ to denote this correlation length. The temporal correlation function, ${\cal D}$ is characterized by a single correlation time, $\talp$. Hence the basic constant parameters of our model are $(\eta_\alpha, \bfV_{\!\!M}, \ell, \talp)$.
   
\section{Evolution equation for the large-scale magnetic field}

\noindent
Here we derive a closed equation for the large-scale magnetic field
by exploiting the homogeneity of the problem in the sheared coordinates
$\bfx$ by working with its conjugate Fourier variable
$\bfk$. Let $\widetilde{Q}(\bfk, t) = \int \mathrm{d}^3x
\exp{(-\mathrm{i}\,\bfk\cendot\bfx)}\,Q(\bfx, t)\,$ be the
Fourier transform of any quantity $Q(\bfx, t)$, with similar
definition in terms of lab-frame coordinates, where $\bfK$ denotes
the conjugate variable to $\bfX$. Note that the lab-frame wavevector $\bfK$ is time-dependent and can be expressed in terms of sheared wavevectors $\bfk$ as, $\bfK(\bfk,\,t)=(k_1-St\,k_2,\,k_2,\,k_3)$; see \Eq{nabsh}.
We need to first solve for $\widetilde{\bfh}(\bfk, t)$ as a functional of $\widetilde{a}(\bfk, t)$ and $\widetilde{\overline{\bfH}}(\bfk, t)$.
This is in general a complicated problem by itself, so for a first attempt we use the standard approach of the first-order smoothing approximation (FOSA)
wherein the term, $\bnabla\cross\left[\,a\bfh-\overline{a\bfh}\right]\,$, is dropped in \Eq{fbsh}. Analogous to Eq.(7.124) of \citet{moffatt1978}, the condition for FOSA to be valid is:
\beq
\frac{\eta_{\alpha}\talp}{\ell^2} \ll 1\,, \qquad\mbox{OR}
\qquad \frac{\eta_{\alpha}}{\eta_T} \ll \frac{\eta_T\,\talp}{\ell^2}\,,
\label{fosa-cond}
\eeq
where we recall that $\ell$ is correlation length of the $\alpha$ fluctuations. The first of this condition comes from the short-correlation
assumption by comparing $\partial \bfh/\partial t$ with
$\bnabla\cross\left[\,a\bfh-\overline{a\bfh}\right]\,$;
$\partial \bfh/\partial t$ is of the order ${\cal O}(h_0/\talp)$ and
$\bnabla\cross\left[\,a\bfh-\overline{a\bfh}\right]\,$ is of the
order ${\cal O}(\alpha_0h_0/\ell)$.
For FOSA to be valid, $\alpha_0h_0/\ell\ll h_0/\talp$.
Using $\alpha_0^2\sim\eta_\alpha/\talp$ (from \Eq{aacorr}), we can
write the first condition in \Eq{fosa-cond}. Similarly, the second
condition comes from comparing $\eta_T\bnabla^2\bfh$ with
$\bnabla\cross\left[\,a\bfh-\overline{a\bfh}\right]\,$  in \Eq{fbsh};
$\eta_T\bnabla^2\bfh$ is of the order ${\cal O}(\eta_T\,h_0/\ell^2)$, and
so, for FOSA to be valid, $\eta_T\,h_0/\ell^2\gg \alpha_0h_0/\ell$,
which yields the second condition in \Eq{fosa-cond} after
rearranging and squaring the terms.

Then the fluctuating magnetic field evolves as:
\beq
\left(\frac{\partial }{\partial t} - \eta_T\bnabla^2\right) \bfh
\;-\; Sh_1\ey \;=\; \bnabla\cross\bfM\,,
\label{fbfosa}
\eeq
where $\bfM(\bfx, t) = a(\bfx, t)\overline{\bfH}(\bfx, t)$ is a
source term for fluctuating magnetic field, and $\bnabla$ is the
time-dependent operator defined in \Eq{nabsh}.
The FOSA solution for the fluctuating magnetic field in the
Fourier space is given by (see \App{app1} for a derivation), 
\beq
\widetilde{\bfh}(\bfk, t) = \int_0^t \mathrm{d}t'\,
\widetilde{G}_{\eta_T}(\bfk,t,t')\left\{\,
\mathrm{i}\bfK(\bfk, t')\cross\widetilde{\bfM}(\bfk, t') +
\ey S(t-t')\left[\mathrm{i}\bfK(\bfk, t')\cross\widetilde{\bfM}(\bfk, t')
\right]_1\,\right\},
\label{hsoln}
\eeq
where the sheared Green's function in Fourier space:
\beq
\widetilde{G}_{\eta_T}(\bfk,t,t') =
\exp\left[-\eta_T\left(k^2(t-t')-S\,k_1\,k_2(t^2-t'^2)+
\frac{S^2}{3}k^2_2(t^3-t'^{\,3})\right)\right]
\label{GF}
\eeq
This is derived in \cite{SS10}. It may be readily verified that the
\Eq{hsoln} satisfies both constraints, $\bfK\cendot\widetilde{\bfh}=0$
and $\widetilde{\bfh}(\bfk, 0) = \bf0\,$. 
By making use of \Eq{hsoln}, and
time-stationary Galilean-Invariant statistics for the $\alpha$ fluctuations
in Fourier space (see \App{app2}),
we obtain the following expression for the mean EMF in Fourier
space, after some straightforward algebra (see \App{app3} for a derivation):
\beq
\widetilde{\overline{\bfE}}(\bfk, t) \;=\; 2\int_0^t \mathrm{d}t'\,
{\cal D}(t-t') \Bigl\{\,\widetilde{\bfU}(\bfk,t, t')
\cross\widetilde{\overline{\bfH}}(\bfk, t')
\; +\; \ey S(t-t')\left[\widetilde{\bfU}(\bfk, t,t')
\cross\widetilde{\overline{\bfH}}(\bfk, t')\right]_1\,\Bigr\}\,,
\label{emfU}
\eeq
where
\beq
\widetilde{\bfU}(\bfk,t, t') \;=\;
\int\frac{\mathrm{d}^3k'}{(2\pi)^3}\,\widetilde{G}_{\eta_T}(\bfk-\bfk',t,t')
\,\mathrm{i}\bfK(\bfk-\bfk', t')
\widetilde{{\cal A}}\left(\bfK(\bfk', t')\right)\,.
\label{Ufou}
\eeq
is a complex velocity field.

Fourier transforming \Eq{lsbsh}, the equation governing
the large-scale field is:
\beq
\frac{\partial \widetilde{\overline{\bfH}}}{\partial t} \;-\;
S\widetilde{\overline{H}_1}\ey \;=\;
\mathrm{i}\bfK(\bfk, t)\cross\widetilde{\overline{\bfE}}
\;-\; \eta_T K^2(\bfk, t)\,\widetilde{\overline{\bfH}}\,,
\qquad \bfK(\bfk, t)\cendot \widetilde{\overline{\bfH}} \;=\; 0\,.
\label{lsbshF}
\eeq
Thus the set of \Eqss{emfU}{lsbshF} describe the evolution of the large-scale
magnetic field, $\widetilde{\overline{\bfH}}(\bfk, t)$ in terms of closed,
linear integro-differential equation, where both shear strength, $S$,
and the $\alpha$-correlation time, $\tau_{\alpha}$, are treated
non-perturbatively. This is the principal general result of this paper,
but solving these in full generality is beyond the scope of the
present work, and we next pursue these equations analytically by making
useful approximations.

\subsection{White-noise $\alpha$ fluctuations}

It is useful to recall basic properties of an exactly solvable limit of
delta-correlated-in-time $\alpha$ fluctuations when the normalized
correlation function ${\cal D}_{\!{\rm WN}}(t) = \delta(t)\,$,
the Dirac delta-function, giving $\talp=0$ from \Eq{corr-time}.
Using this in \Eqs{emfU}{Ufou}, and noting that
$\widetilde{G}_{\eta_T}(\bfk-\bfk',t,t)=1$ from \Eq{GF}, we
find the mean EMF:
\beq
\widetilde{\overline{\bfE}}_{\rm WN}(\bfk, t) =
\widetilde{\bfU}_{\!\rm WN}(\bfk,t)\cross
\widetilde{\overline{\bfH}}(\bfk, t)\quad\mbox{with}\quad
\widetilde{\bfU}_{\!\rm WN}(\bfk,t)=\mathrm{i}\bfK(\bfk, t)\eta_{\alpha}
+ \bfV_{\!\!M}\,,
\label{wnemfF}
\eeq
where the $\alpha$-diffusivity, $\eta_{\alpha}={\cal A}({\bf 0})$, is given in \Eq{aacorr} and the Moffatt drift velocity
$\bfV_{\!\!M}$ is defined in \Eq{kmcondef}. The \emph{Kraichnan diffusivity}, $\eta_K$, is defined as, $\eta_K=\eta_T-\eta_\alpha$. Using these in \Eq{lsbshF} leads to the solution for the large-scale magnetic field (see \cite{SS14} for more details):
\beq
\widetilde{\overline{\bfH}}(\bfk, t) \;=\;
\widetilde{{\cal G}}(\bfk, t)\,\left[\,
\widetilde{\overline{\bfH}}(\bfk, 0)+\ey St\,\widetilde{\overline{H}}_1(\bfk, 0)\,\right]\,,\qquad\qquad\bfk\cendot
\widetilde{\overline{\bfH}}(\bfk, 0) \;=\; 0\,.
\label{Hsoln-wn}
\eeq
where
\begin{eqnarray}
\widetilde{{\cal G}}(\bfk, t) &\;=\;& \exp{\left\{\,-\int_0^t\,\mathrm{d}t' 
\left[\,\eta_K K^2(\bfk, t')\,+\,\mathrm{i}\, \bfV_{\!\!M}\cendot\bfK(\bfk, t')\,\right]\,\right\}}\nonumber\\[1em]
&\;=\;& \exp{\Bigl\{-\eta_K\left[\,k^2t - Sk_1k_2t^2 + (S^2/3)k_2^2t^3\,\right] \,-\, \mathrm{i}\left[\,\left(\bfV_{\!\!M}\cendot\bfk\right) t - (S/2) V_{\!M1}k_2 t^2\,\right]\Bigr\}}\,,
\nonumber\\
&&\label{greenfn}
\end{eqnarray}
This solution is identical to the one obtained in \cite{SS14}. Thus
we find that the inclusion of the turbulent diffusion term in
determining the mean EMF makes no difference for the dynamo solution
in the white-noise limit. In agreement with earlier findings
\citep{Kra76,moffatt1978,SS14}, we see from above that
the $\alpha$-diffusivity causes a reduction in the turbulent diffusion of
the fields, and if it is sufficiently strong, i.e., when $\eta_K<0$,
this can lead to an instability giving growth of large-scale magnetic field.
Also, the Moffatt drift does not couple to the dynamo growth/decay
and contributes only to the phase.

\section{Axisymmetric large-scale dynamo equation with finite $\talp$}

\noindent
We now turn to the principal aim of this work where we are more interested
in exploring the possibility of large-scale dynamo even when the
$\alpha$ fluctuations are weak, i.e., when $\eta_K>0$, by taking the
memory effects into account. Assuming small but finite correlation time
for $\alpha$ fluctuations, $\talp \neq 0$, we reduce the general
set of \EEqss{emfU}{lsbshF} into a partial differential equation
governing the dynamics of large-scale magnetic field which evolves
over times much larger than $\talp$.
In this case, the normalized time correlation function, ${\cal D}(t)$,
is significant only for times $t\leq \talp$ and it becomes negligible
for larger times. The generalized mean EMF as given in \Eq{emfU} involves
a time integral which can be solved under the small $\talp$ approximation.

Since the limit $\lim\limits_{\tau_{\alpha}\to 0}
\widetilde{\overline{\bfE}}(\bfk, t) =
\widetilde{\overline{\bfE}}_{\rm WN}(\bfk, t)$, given by \Eq{wnemfF},
is non-singular, we proceed by making the following \emph{ansatz} where,
for small $\talp$, the mean EMF can be expanded in a power series in
$\talp$ as:
\beq
\widetilde{\overline{\bfE}}(\bfk, t) \;=\;
\widetilde{\overline{\bfE}}_{\rm WN}(\bfk, t)
\,+\, \widetilde{\overline{\bfE}}^{(1)}(\bfk, t) \,+\,
\widetilde{\overline{\bfE}}^{(2)}(\bfk, t) \,+\, \,\dots
\label{emfexp}
\eeq
where $\widetilde{\overline{\bfE}}_{\rm WN}(\bfk, t) \sim {\cal O}(1)$
and $\widetilde{\overline{\bfE}}^{(n)}(\bfk, t) \sim
{\cal O}(\tau_{\alpha}^n)$ for $n \geq 1$.
Below we verify this ansatz up to $n= 1$, for slowly varying magnetic
fields. From \Eq{emfU} we determine
$\widetilde{\overline{\bfE}}(\bfk, t)$ to first
order in $\tau_{\alpha}\,$, for $t\gg\tau_{\alpha}\,$, by
(i) changing the integration variable from $t'$ to $s=t-t'$;
(ii) setting the upper limit of the time integral to $+\infty$,
since ${\cal D}(s)$ is significant only for times $s\leq\talp$ as
mentioned above, suggesting that only short times $0\leq s < \talp$
contribute appreciably to the integral in \Eq{emfU};
and (iii) keeping the terms inside the $\{\;\}$ in the integrand
of \Eq{emfU} up to only first order in $s$. To be able to expand in $s$, we need to first express the \Eq{Ufou} in lab frame wave vector $\bfK = \bfK(\bfk,t')=\bfK(\bfk,t-s)$, so that the green's function in \Eq{GF} and therefore the complex velocity field $\widetilde{\bfU}$ in \Eq{Ufou} becomes time-translational symmetric \footnote{Greens's function in \Eq{GF} when expressed in lab frame wave vector becomes time-translational symmetric, i.e.,\\ 
$ \widetilde{G}_{\eta_T}(\bfk,t,t')  = \widetilde{G}_{\eta_T}(\bfK(\bfk,t'), t-t',0) 
  =\exp(-\eta_T\{K^2(t-t')-S\,K_1\,K_2(t-t')^2+\frac{S^2}{3}K^2_2(t-t')^3\})
$.
}. 
 
We first rewrite the mean EMF, given in \Eq{emfU}, as
\beq
\widetilde{\overline{\bfE}}(\bfk, t) \;=\; 2\int_0^{\infty} \mathrm{d}s\,{\cal D}(s)\left\{\,\widetilde{\bfU}(\bfK, s)\cross\widetilde{\overline{\bfH}}(\bfk, t-s) \;+\; 
\ey Ss\left[\widetilde{\bfU}(\bfK, s)\cross\widetilde{\overline{\bfH}}(\bfk, t-s)
\right]_1\,\right\}\,,
\label{emfU2}
\eeq
where the complex velocity field, $\widetilde{\bfU}$, is
\beq
\widetilde{\bfU}(\bfK, s) \;=\; 
\int\frac{\mathrm{d}^3K'}{(2\pi)^3}\,\widetilde{G}_{\eta_T}(\bfK-\bfK',s)\,\mathrm{i}(\bfK-\bfK')
\widetilde{{\cal A}}\left(\bfK'\right)\,.
\label{Ufou2}
\eeq

\noindent
Equation~(\ref{Ufou2}) is obtained by changing the integration variable in (\ref{Ufou}) to $\bfK' =\bfK(\bfk', t') = \left(k'_1 - S(t-s)k'_2\,, k'_2\,, k'_3\right)\,$ --- which has unit Jacobian giving $\mathrm{d}^3k'=\mathrm{d}^3K'$.

We make further simplification by considering only axisymmetric modes
for which $k_2=0$. Note that for the non-axisymmetric modes,
$\bfK(\bfk, t)=\ex(k_1 - St\,k_2) + \ey k_2 + \ez k_3$ increases
monotonically with time, increasing the wavenumber, which would eventually decay by turbulent diffusivity. 
Therefore we focus our attention only on axisymmetric modes, for
which $\bfK(\bfk, t-s)=\bfK(\bfk, t)=\bfk=(k_1, 0, k_3)\,$.

Let us first work out $\widetilde{\bfU}(\bfk, s)$ and
$\widetilde{\overline{\bfH}}(\bfk, t-s)$ correct up to ${\cal O}(s)$. \\

\noindent
$\bullet\;\;\underline{\widetilde{\bfU}(\bfk, s)\;
\mbox{to}\; {\cal O}(s) :}$
Taylor expanding $\widetilde{\bfU}(\bfk, s)$ gives,
\begin{eqnarray}
\widetilde{\bfU}(\bfk, s)& \;=\;& \widetilde{\bfU}(\bfk, 0) \,+\,
s \frac{\partial \widetilde{\bfU}}{\partial s}{\Biggl {\vert}}_{s = 0}
\,+\, {\cal O}(s^2)\,.
\\[2ex]
 \mbox{where}\qquad \widetilde{\bfU}(\bfk, 0)  &\;=\;&
 \mathrm{i}\bfk\, \eta_{\alpha} \;+\; \bfV_{\!\!M}
 \qquad\mbox{(from \Eq{wnemfF})}\,,
 \label{Pdef}\\[2ex]
\mbox{and}\quad\frac{\partial \widetilde{\bfU}}{\partial s}
 {\Biggl {\vert}}_{s = 0} &\;=\;& -\mathrm{i}\, \eta_T
 \int \frac{\mathrm{d}^3K'}{(2\pi)^3}\, (\bfK-\bfK')^2 (\bfK-\bfK')
 \widetilde{{\cal A}}(\bfK')\,.\label{Qdef}
\end{eqnarray}
\Eq{Qdef} is obtained by differentiating \Eq{Ufou2} w.r.t $s$ and taking the limit $s\to 0$, note that $\bfK = \bfk$, since $k_2=0$. Using the Fourier transform for $\widetilde{{\cal A}}$, together with the properties of delta-function, we get  

\begin{eqnarray}
\frac{\partial \widetilde{\bfU}}{\partial s}
 {\Biggl {\vert}}_{s = 0} && =  -\mathrm{i}\,\eta_T \Biggl\{ k^2\,\biggl(\bfk {\cal A}(\bfxi) + \mathrm{i}\, [\bnabla{\cal A}(\bfxi)]\biggl) + 2\,\mathrm{i}\,\bfk\, \left(\bfk{\bf\cdot}[\bnabla{\cal A}(\bfxi)]\right)\nonumber \\
&& - \bfk [\bnabla^2{\cal A}(\bfxi)] -\mathrm{i}\,[\bnabla^2\{\bnabla{\cal A}(\bfxi)\}]
 -2(\bfk\cdot\bnabla)\{\bnabla{\cal A}(\bfxi)\} \biggr\}_{{\small\bfxi}=0}
 \label{Udiff}
\end{eqnarray}

Equation~(\ref{Udiff}) can be evaluated once we know the functional form for spatial correlator ${\cal A}(\bfxi)$. Neglecting derivatives
of ${\cal A}$ that are higher than the first order --- see \cite{S16} for detail --- we have: 
 
\beq
\frac{\partial \widetilde{\bfU}}{\partial s}
 {\Biggl {\vert}}_{s = 0} \,=\, -\eta_T k^2 \left(\mathrm{i}\bfk \eta_{\alpha}
\,+\, \bfV_{\!\!M}\right) \,-\,
2\eta_T (\bfk\,\cendot \bfV_{\!\!M}) \bfk\,.
\label{Qsimple}
\eeq
\Eqs{Pdef}{Qsimple} together, thus provides the function
$\widetilde{\bfU}(\bfk, s)$ correct up to ${\cal O}(s)$.\\

\noindent
$\bullet\;\;\underline{\widetilde{\overline{\bfH}}(\bfk, t-s)\;
\mbox{to}\; {\cal O}(s) :}$ We write as,
\beq
\widetilde{\overline{\bfH}}(\bfk, t-s) \;=\;
\widetilde{\overline{\bfH}}(\bfk, t) \,-\,s 
\frac{\partial \widetilde{\overline{\bfH}}(\bfk, t)}{\partial t}
\,+\, \ldots\,.
\label{Hexp}
\eeq
where it is assumed that $\bigl{\vert} \widetilde{\overline{\bfH}}
\bigr{\vert} \,\gg\, s\bigl{\vert}\partial
\widetilde{\overline{\bfH}}/\partial t\bigr{\vert} \,\gg\;
s^2\bigl{\vert}\partial^2 \widetilde{\overline{\bfH}}/\partial t^2
\bigr{\vert}\,,\,\mbox{etc}\,$.
In \Eq{Hexp}, we need $\partial \widetilde{\overline{\bfH}}/\partial t$
only up to ${\cal O}(1)$ to find $\widetilde{\overline{\bfH}}(\bfk, t-s)$
up to ${\cal O}(s)$.
We write this by substituting \Eq{wnemfF} in \Eq{lsbshF} and using
$\widetilde{\bfU}_{\!\rm WN}(\bfk,t)=\widetilde{\bfU}_{\!\rm WN}(\bfk)=
\mathrm{i}\bfk\eta_{\alpha} + \bfV_{\!\!M}$:
\beq
\frac{\partial \widetilde{\overline{\bfH}}}{\partial t}
\biggl\vert_{{\cal O}(1)}  \;=\;
S\widetilde{\overline{H}_1}\ey \;+\;
\mathrm{i}\bfk\cross\widetilde{\overline{\bfE}}_{\rm WN}
\;-\; \eta_T k^2\,\widetilde{\overline{\bfH}}=\;
S\widetilde{\overline{H}_1}\ey \;-\; \left(\eta_{K}\, k^2\,+\,
\mathrm{i}\,\bfk\cdot\bfV_{\!\!\rm M}\right)\widetilde{\overline{\bfH}}
\label{Hdot-0thord}
\eeq
Time-integral in \Eq{emfU2} is then solved by using definitions provided
in \Eqs{aacorr}{corr-time} when we substitute the expressions derived
just above for the terms in $\{\;\}$ in \Eq{emfU2}. We get after straightforward algebra the following
expression for the mean EMF which is correct upto ${\cal O}(\talp)$:
\begin{eqnarray}
\widetilde{\overline{\bfE}}(\bfk, t) \;&=&\;
\widetilde{\overline{\bfE}}_{\rm WN}(\bfk, t)\,+\,
\talp\biggl\{(\mathrm{i}\,\bfk\cdot\widetilde{\bfU}_{\rm WN})\,
\widetilde{\overline{\bfE}}_{\rm WN}\,-\,
2 \eta_{\rm T}\,(\bfk\cdot\bfV_{\!\!\rm M})\,
\bfk\cross\widetilde{\overline{\bfH}}\biggl\}\,+\nonumber\\[1ex]
&&+\,S\,\talp\biggl\{\widetilde{\overline{H_1}}\,\ey\cross
\widetilde{\bfU}_{\rm WN}\,+\,
\ey\left[\widetilde{\bfU}_{\!\rm WN}\cross\widetilde{\overline{\bfH}}\right]_1
\biggl\}
\label{emfotau}
\end{eqnarray}
This verifies the ansatz of \Eq{emfexp} up to $n=1$, as claimed.
It is important to note that the \Eq{emfotau} is valid only for slowly
varying large--scale magnetic fields. To lowest order this condition can
be explicitly stated as: $\,\bigl{\vert} \widetilde{\overline{\bfH}}
\bigr{\vert} \gg \tau_{\alpha}\bigl{\vert}\partial
\widetilde{\overline{\bfH}}/\partial t \bigr{\vert}\,$.
To obtain the sufficient condition for the validity of
\Eq{emfotau}, use \Eq{Hdot-0thord} for
$\,\partial \widetilde{\overline{\bfH}}/\partial t\,$ to get the following conditions for three dimensionless quantities which need to be small:
\beq
\vert S\tau_{\alpha}\vert \,\ll\, 1\,,\qquad\qquad
\vert \eta_K k^2 \tau_{\alpha}\vert \,\ll\, 1\,,\qquad\qquad
\vert k V_M \tau_{\alpha}\vert \,\ll\,1\,.
\label{cond}
\eeq
Since we have expanded EMF in small $\talp$, it is only the
first of the two FOSA conditions in \Eq{fosa-cond} that becomes
relevant. This must be added to the above three conditions for
\Eq{emfotau} to be valid.
Using \Eq{emfotau} in \Eq{lsbshF} we obtain:
\begin{eqnarray}
\frac{\partial \widetilde{\overline{\bfH}}}{\partial t} &=&
\left[\,S\widetilde{\overline{H}}_1\ey \,+\, \eta_{\alpha} k^2
\widetilde{\overline{\bfH}} \,-\,\mathrm{i}\,(\bfk\cendot\bfV_{\!\!M})
\widetilde{\overline{\bfH}}\,\right]
\left[\,1 \,+\, \mathrm{i}\,(\bfk\cendot\bfV_{\!\!M})\tau_{\alpha} \,-\,
\eta_{\alpha} k^2\tau_{\alpha}\,\right] 
\,-\,\eta_T\,k^2 \widetilde{\overline{\bfH}} + \nonumber\\[1em]
&& +\,2\,{\rm i}\,\eta_T k^2\tau_\alpha(\bfk\cdot\bfV_{\! \! M})
\widetilde{\overline{\bfH}}+S\tau_{\alpha}
\left[\,V_{M2} \widetilde{\overline{H}}_3 -
V_{M3} \widetilde{\overline{H}}_2 -
\mathrm{i}\,\eta_{\alpha} k_3 \widetilde{\overline{H}}_2\,\right]
\left[\,-\mathrm{i}\,k_3\ex + \mathrm{i}\,k_1 \ez\,\right]\,,
\nonumber\\[1em]
&&\mbox{with}\quad \bfk \;=\; (k_1\,, 0\,, k_3)\,,\qquad \mbox{and}\qquad
k_1\widetilde{\overline{H}}_1 \,+\, k_3\widetilde{\overline{H}}_3 \;=\; 0\,.
\label{lsb-axisym}
\end{eqnarray}

\emph{\EEq{lsb-axisym} is the linear partial differential equation
obtained by reducing the linear integro-differential equation (see \Eqss{emfU}{lsbshF}) under the condition of (\ref{cond}). Nonetheless, it describes the evolution of an axisymmetric, large--scale
magnetic field over times that are much larger
than $\talp\,$. It depends on (i) the diffusivity $\eta_T$; (ii) properties of alpha-correlation in terms of $\eta_\alpha$, $\bfV_{\!\!M}$ and $\talp$; (iii) shear $S$.  These must satisfy the three conditions given in \Eq{cond} and first condition in \Eq{fosa-cond} for the validity of the \Eq{lsb-axisym}. We note here again that the set of \Eqss{emfU}{lsbshF} are non-perturbative in both $S$ and $\talp$, whereas \Eq{lsb-axisym} is valid only when $\vert S\tau_{\alpha}\vert \,\ll\, 1\,$}.

\section{Growth rate of modes when $\talp$ is non-zero}

As usual in numerical works on the related subject
\citep[see, e.g.,][]{BRRK08,SJ15} where ``horizontal''
(plane of shear; in this case the $X_1-X_2$ plane) averages are performed to
define the large--scale magnetic fields, it is therefore useful to consider
one--dimensional propagating modes. This is equivalent to setting $K_1$ and
$K_2$ equal to zero. Here we only need to set $k_1=0$ in \Eq{lsb-axisym}.
In this case the wavevector $\bfk = (0,0,k)$ points along the
``vertical'' ($\pm\ez$) direction, thus resulting in a uniform
$\widetilde{\overline{H}}_3$ which is of no interest for dynamo action.
Hence we set $\widetilde{\overline{H}}_3 = 0$, and take
$\widetilde{\overline{\bfH}}(k, t) = \widetilde{\overline{H}}_1(k, t)\ex
\,+\, \widetilde{\overline{H}}_2(k, t)\ey\,$. Making these substitutions
in \Eq{lsb-axisym} we find:
\begin{eqnarray}
\frac{\partial \widetilde{\overline{\bfH}}}{\partial t} &\;=\;&
\left[\,S\widetilde{\overline{H}}_1\ey \,+\, \eta_\alpha k^2
\widetilde{\overline{\bfH}} \,-\,\mathrm{i}\,kV_{M3}
\widetilde{\overline{\bfH}}\,\right]
\left[\,1 \,+\, \mathrm{i}\,kV_{M3}\tau_{\alpha} \,-\,
\eta_{\alpha} k^2\tau_{\alpha}\,\right]
\,-\,\eta_T\,k^2 \widetilde{\overline{\bfH}}\,+\nonumber \\[1em] 
&&+2\,{\rm i}\,\eta_T\,\tau_\alpha(k^3 V_{M3})\widetilde{\overline{\bfH}}
\,+\,S\left[\,\mathrm{i}\,kV_{M3}\tau_{\alpha} \,-\,
\eta_{\alpha} k^2\tau_{\alpha}\,\right] 
\widetilde{\overline{H}}_2\ex
\label{lsb1dF}
\end{eqnarray}
Seeking modal solutions of the form,
\beq
\widetilde{\overline{\bfH}}(k, t) \;=\;
\left[\widetilde{\overline{H}}_{01}(k)\ex \,+\,
\widetilde{\overline{H}}_{02}(k)\ey\right]\,\exp{(\lambda t)}\,,
\label{modal}
\eeq
and substituting this in \Eq{lsb1dF} we get the following
dispersion relation:
\begin{eqnarray}
\lambda_{\pm} &\;=\;&
\,-\, \eta_K k^2 \,-\, \eta_{\alpha}^2 k^4\tau_{\alpha}
\,+\, (k V_{\!M3})^2 \tau_{\alpha} \,+\,
\mathrm{i}\,k V_{\!M3}
\left[\,2(\eta_{\alpha}+\eta_T) k^2 \tau_{\alpha} \,-\, 1\,\right] 
\nonumber\\[1em]
&&\qquad\quad \pm\;\vert S\vert\,\sqrt{
\left[\mathrm{i}kV_{M3}\tau_\alpha - \eta_\alpha k^2\tau_\alpha\right]
\left[\,1 \,+\, \mathrm{i}\,kV_{M3}\tau_{\alpha} \,-\,
\eta_{\alpha} k^2\tau_{\alpha}\,\right]\;}
\label{disprel}
\end{eqnarray}
We are more interested in the growth rate $\gamma = {\rm Re}\{\lambda\}\,$,
as the dynamo action corresponds to the case when $\gamma > 0\,$. From
the dispersion relation~(\ref{disprel}) we have:
\begin{eqnarray}
\gamma_{\pm} &\;=\;& {\rm Re}\{\lambda_{\pm}\} \;=\;-\, \eta_K k^2 \,-\, \eta_{\alpha}^2 k^4\tau_{\alpha}
\,+\, (k V_{\!M3})^2 \tau_{\alpha}
\;\pm\; \vert S\vert\left[\chi_R^2 \,+\, \chi_I^2\right]^{1/4}\,\cos{(\psi/2)}\,,
\nonumber\\[1em]
\mbox{where}&&\quad \cos{(\psi)} \;=\; \frac{\;\chi_R\;}{(\chi_R^2+\chi_I^2)^{1/2}}\,,\qquad \cos(\psi/2)=\frac{\sqrt{\chi_R+\sqrt{\chi_R^2+\chi_I^2}}}{\sqrt{2}\left(\chi_R^2 \,+\, \chi_I^2\right)^{1/4}}\,,
\nonumber\\[1em]
\chi_R &\;=\;& \eta_\alpha k^2\tau_\alpha \left(\,\eta_{\alpha}
k^2 \tau_{\alpha} - 1\right) \;-\; \left(kV_{\!M3}\tau_{\alpha}\right)^2
\,,\quad
\chi_I = -kV_{M3}\tau_{\alpha} \left(\,2\eta_{\alpha} k^2 \tau_{\alpha}
\,-\, 1\right) .
\label{gamma}
\end{eqnarray}
Below we make some comments about the growth rate derived above:
\begin{itemize}
\item The growth rate $\gamma$ of the large-scale dynamo is linear in
the shear rate $|S|$, assuming that the parameters
($\eta_K,\,\eta_\alpha,\,V_{\!M3},\,\talp$) are all independent of $S$.
This linear scaling is observed in earlier numerical works
\citep{BRRK08,You08a,You08b,SJ15}.
\item For zero shear, the growth rate as given from \Eq{gamma} becomes
identical to the one derived in \cite{S16}, where the generalization to
the Kraichnan problem as well as the possibility of Moffatt drift driven
dynamos were explored in detail.
\item The last term involving shear in \Eq{gamma} is identical to the
corresponding term in the expression for the growth rate derived in
\cite{SS14}, with an important difference being that there the angle
$\psi$ was defined using tangent function, which introduces error when
either of the two, $\chi_R$ and $\chi_I$, take negative values. Here we
correct this by explicitly writing $\cos{(\psi/2)}$ in terms of
$\chi_R$ and $\chi_I$.
\end{itemize}
\subsection{Dimensionless growth rate function}

\begin{figure}
\centering
\includegraphics[width=0.49\columnwidth]{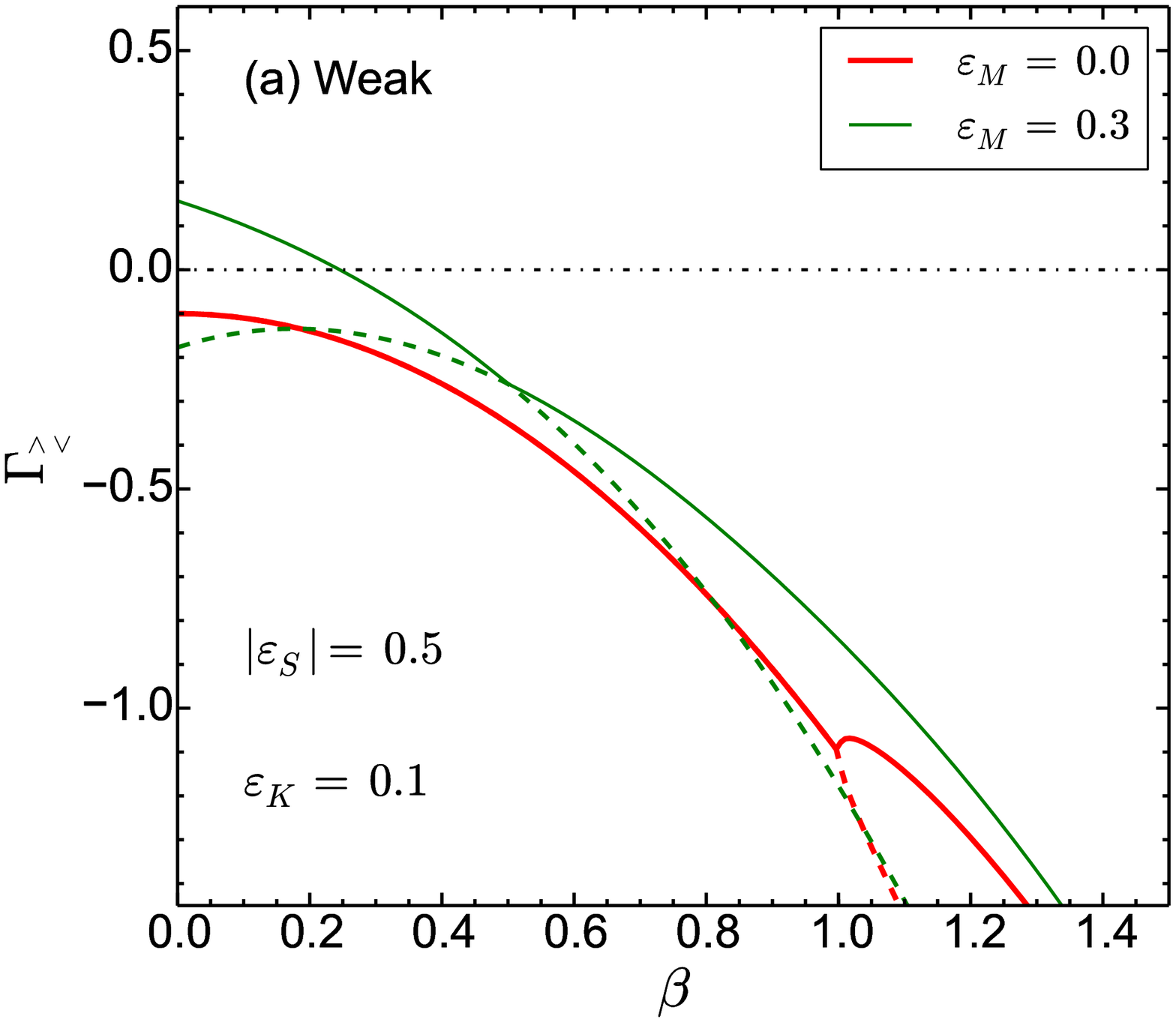}
\includegraphics[width=0.49\columnwidth]{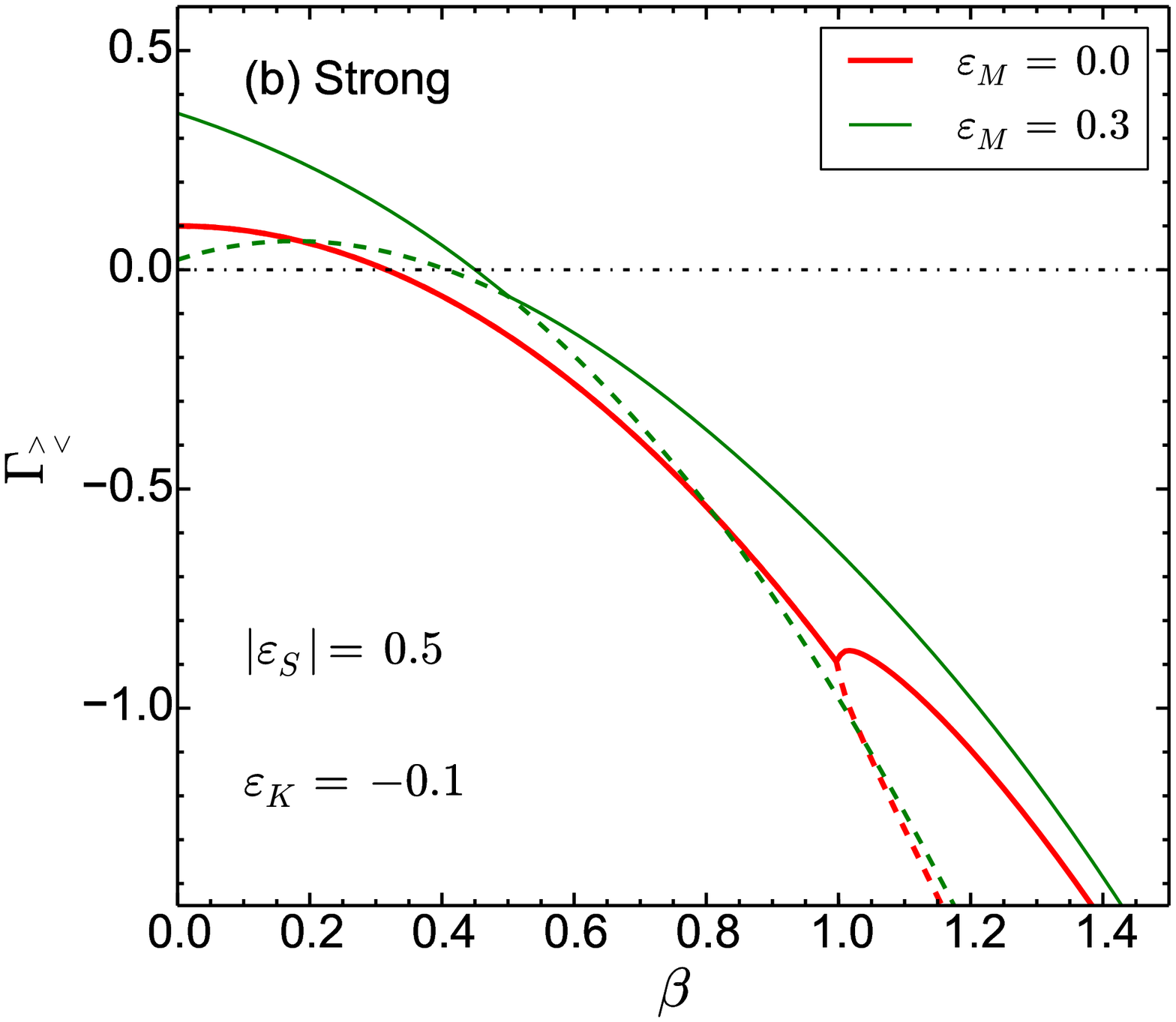}
\caption{
The two roots, $\Gamma_>$ (solid) and $\Gamma_<$ (dashed), of
the growth rate function defined in \Eq{Gamma}
are shown as function of $\beta$ for $\varepsilon_M=0$ (red; thick)
and $0.3$ (green; thin) with $|\varepsilon_S|=0.5$, where left
and right panels correspond to weak ($\varepsilon_K=0.1$)
and strong ($\varepsilon_K=-0.1$) $\alpha$ fluctuations, respectively.
}
\label{fig1}
\end{figure}

The growth rate function $\Gamma$ is defined using dimensionless
quantities,
\beq
\Gamma_{\pm} = \gamma_{\pm}\tau_\alpha\,;\quad\;
\beta = \eta_\alpha k^2\tau_\alpha\,;\quad\;
\varepsilon_S = S\tau_\alpha\,;\quad\;
\varepsilon_K = \eta_Kk^2\tau_\alpha\,;\quad\;
\varepsilon_M = kV_{M3}\tau_\alpha\,,
\label{dimless}
\eeq
where $\beta$ and $\varepsilon_K$ measure the wavenumber
of modal mean-magnetic field in terms of $\eta_\alpha$ and $\eta_K$,
respectively. With the first condition of \Eq{fosa-cond},
$\beta/k^2=\eta_\alpha\talp\ll\ell^2$. These parameters can vary as,
\beq
0 \;\leq\; \beta \;\ll\; (k\ell)^2\,; \quad\;
\beta + \varepsilon_K \;>\; 0\,;\quad\;
\vert\varepsilon_S\vert \;\ll\; 1\,;\quad\;
\vert\varepsilon_K\vert \;\ll\; 1\,;\quad\;
\vert\varepsilon_M\vert \;\ll\; 1\,.
\label{parrange}
\eeq 
The parameter $\beta$ can be larger or smaller than unity
depending on whether the mean-field varies over scales smaller
or larger than $\ell$, respectively. The second condition comes
from $\beta + \varepsilon_K =\eta_T\,k^2\talp> 0$, and last three
constraints come from \Eq{cond}.
Multiplying the expression for $\gamma_\pm$ in \Eq{gamma} by $\talp$,
and denoting by $\Gamma_>$ ($\Gamma_<$)
larger (smaller) of $\Gamma_+$ and $\Gamma_-$, we get
\beq
\Gamma_{\stackrel{>}{<}} \,= -\,\varepsilon_K\,-\,\beta^2 \,+\, \varepsilon_M^2
\,\pm\, \frac{\vert\varepsilon_S\vert}{\sqrt{2}}\;
\sqrt{\beta(\beta - 1)-\varepsilon_M^2 \,+\,
\sqrt{\bigl[\beta(\beta - 1)+\varepsilon_M^2\bigr]^2+\varepsilon_M^2}}\,.
\label{Gamma}
\eeq
Note that the radicand in \Eq{Gamma} is greater than zero.
In \Fig{fig1} we show the behaviour of $\Gamma$ as function of $\beta$
by keeping other parameters as fixed. Below we list some properties of
the growth rate function as defined in \Eq{Gamma}:

\begin{enumerate}
\item [1.] For fixed $\varepsilon_K$, $\beta$ and $\varepsilon_M$,
$\Gamma_>$ ($\Gamma_<$) increases (decreases) monotonically with shear.
\item [2.] When $\varepsilon_M$ is non-zero, then the radicand in \Eq{Gamma}
vanishes at $\beta=1/2\,$, where the two roots coincide; see green solid and
dashed curves in \Fig{fig1}. Both roots are identical for
$0 \leq \beta \leq 1$ when $\varepsilon_M=0$ and branch out for $\beta>1$; see red solid and dashed curves.
\item [3.] In the absence of the Moffatt drift, the necessary condition for
dynamo action is that the $\alpha$ fluctuations must be strong, i.e.,
$\varepsilon_K<0$, regardless of the strength of the shear
parameter $|\varepsilon_S|$ which should
be kept smaller than unity in the present model. The dynamo is then driven
through the $-\varepsilon_K$ term in \Eq{Gamma} by the process
of negative diffusion first suggested by \cite{Kra76}.
\item [4.] Moffatt drift always contributes positively to the dynamo growth.
Considering the case of zero shear, we
see from \Eq{Gamma} that $\varepsilon_M > \varepsilon_M^{\rm crit}$,
with $\varepsilon_M^{\rm crit} = \sqrt{\varepsilon_K + \beta^2}$, can always
facilitate LSD in both, weak and strong $\alpha$ fluctuation, regimes,
for sufficiently low values of $\beta$ such that
$\varepsilon_M^{\rm crit} \ll 1$.
\item [5.] The growth rate is always negative for $\beta \gg 1$ due to the
$-\beta^2$ term in \Eq{Gamma} as is also shown in \Fig{fig1}. 
\item[6.]The growth rate for $\beta\approx 0$ i.e, largest scale possible is given for small values of $\varepsilon_M \ll 1$ as,
$$\Gamma_{>} \simeq -\varepsilon_K + \frac{|\varepsilon_S|}{\sqrt{2}}|\varepsilon_M|^{1/2} $$ which implies moffatt drift couples strongly with shear and growth is possible for weak $\alpha$-fluctuations i.e., $\varepsilon_K>0$ when $|\varepsilon_S||\varepsilon_M|^{1/2}>\sqrt{2}\,\varepsilon_K$. 

\end{enumerate}

\subsection{Growth rates as functions of the wavenumber}

We henceforth consider only the dominant root $\Gamma_>$ and study its
wavenumber dependence. Following SS14, we first identify natural
length and time scales whose corresponding wavenumber and frequency
are defined as,
\beq
k_\alpha\;=\;(\eta_\alpha \talp)^{-1/2} \;>\;0\,;
\qquad \sigma = \vert \eta_K \vert k^2_\alpha\,.
\label{kalp}
\eeq
where $k_\alpha$ can be recognized as inverse diffusion length due to 
$\alpha-$ diffusivity $\eta_\alpha$. Here $|k|>k_\alpha$ and $|k|<k_\alpha$ are called high and
low wavenumbers, respectively.
From \Eqs{dimless}{kalp} we see that $\beta=(k/k_\alpha)^2$.
Since the parameters $\varepsilon_K$ and $\varepsilon_M$ involve
wavenumber $k$ in their definitions, we find it better to rewrite
an expression for $\Gamma_>$ using new dimensionless
\emph{dynamo numbers}, which are defined in terms of known
constants:
\beq
{\cal D}_\alpha \,=\, \frac{\eta_\alpha}{\eta_T}\;,\qquad\qquad
{\cal D}_M \,=\, \frac{V_{\!M3}^2 \talp}{\eta_\alpha}
\label{dynnos}
\eeq

We first make use of \Eqss{gamma}{dynnos} to express the growth rates as function of wavenumbers and constant dynamo parameters 
in the regime of weak and strong, regimes of $\alpha$--fluctuations. \\

\begin{figure}
\centering
\includegraphics[width=0.49\columnwidth]{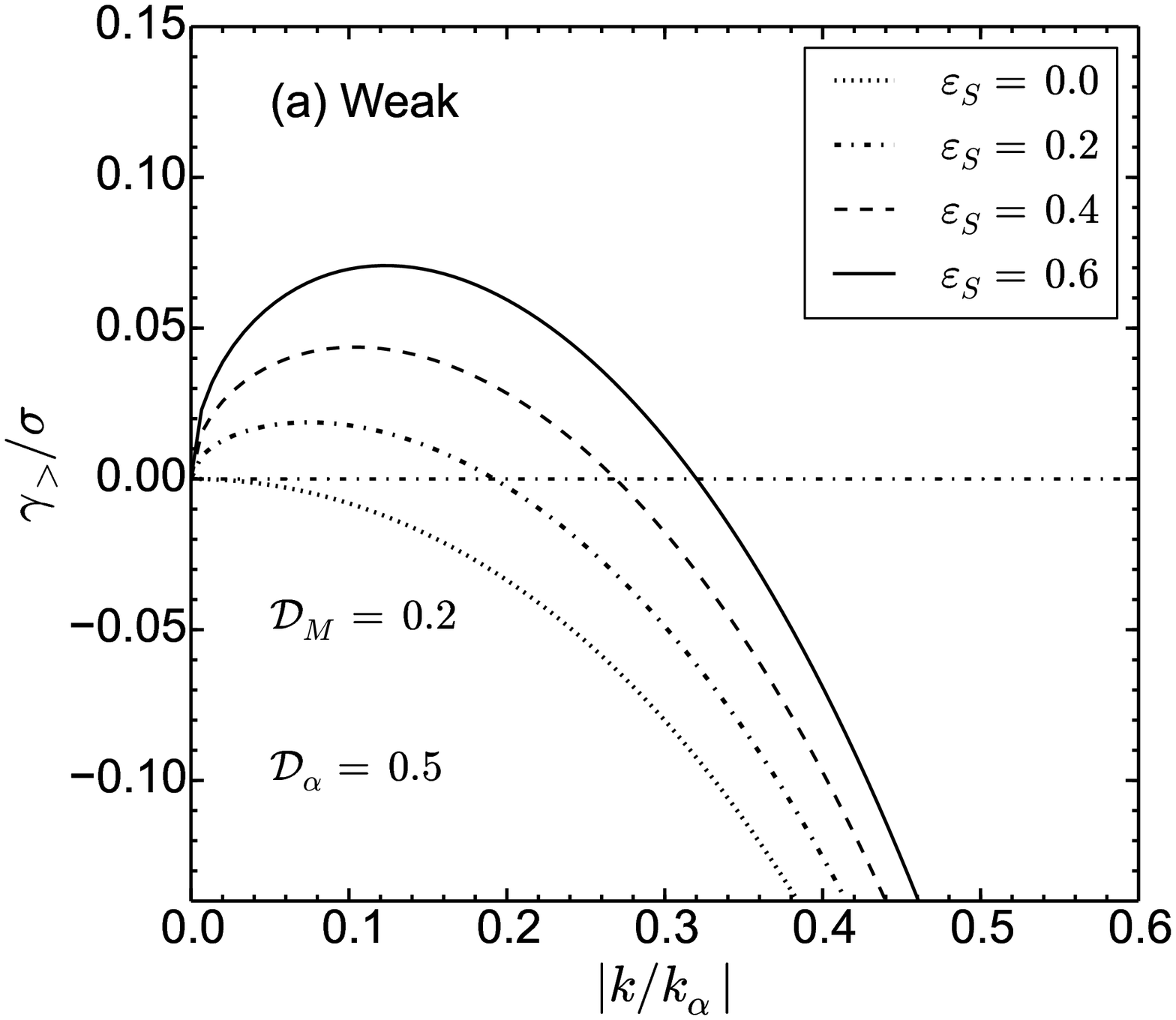}
\includegraphics[width=0.49\columnwidth]{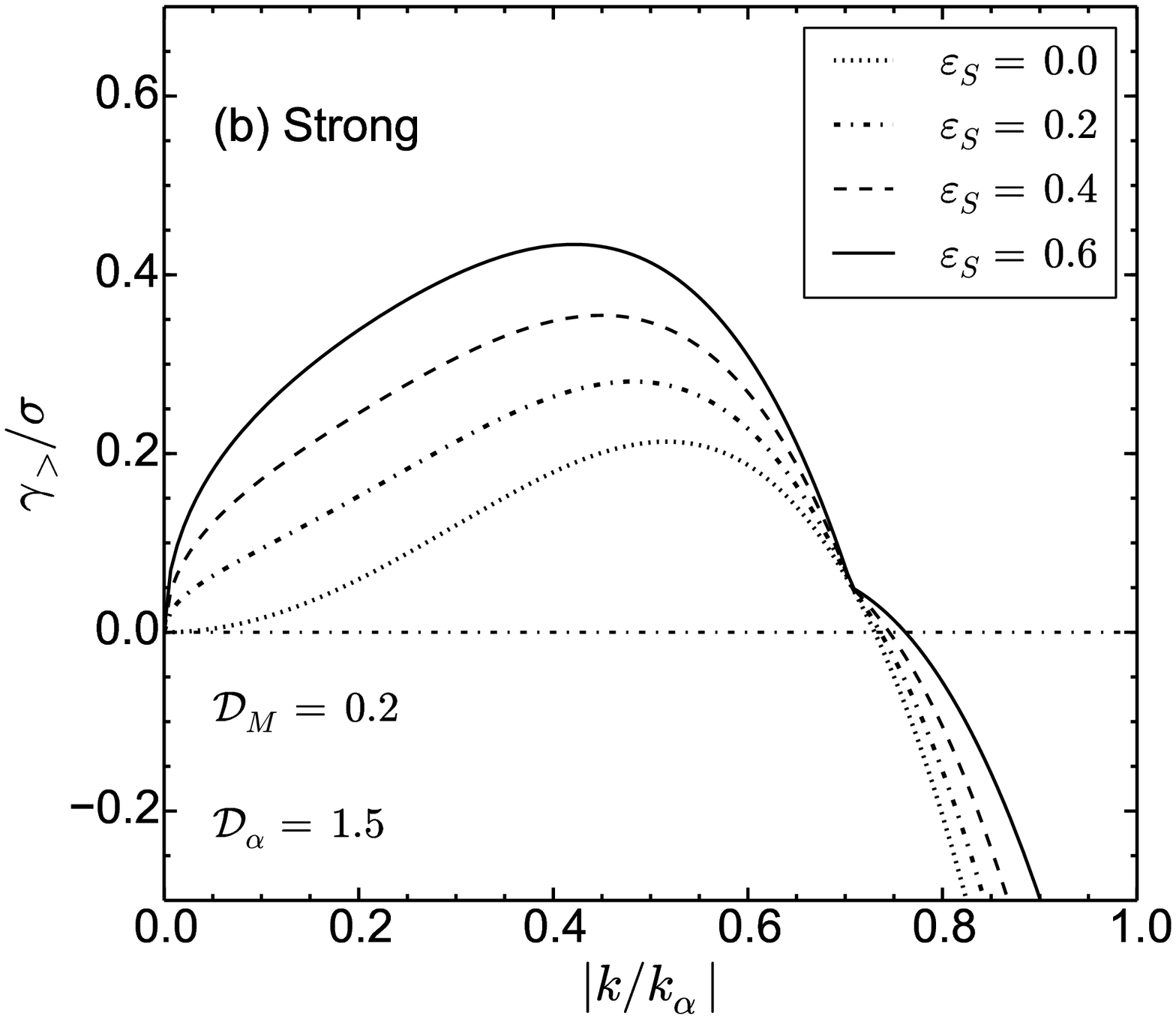}
\caption{
Normalized growth rate $\gamma_>/\sigma$ as function of $|k/k_\alpha|$
for ${\cal D}_M=0.2$. Left and right panels correspond to weak
(${\cal D}_\alpha=0.5$) and strong (${\cal D}_\alpha=1.5$) $\alpha$ fluctuations,
respectively. Solid, dashed, dash-dotted and dotted curves correspond to
$|\varepsilon_S|=0.6$, 0.4, 0.2, and 0, respectively.
}
\label{fig2}
\end{figure}

\paragraph{{\bf Weak $\alpha$ fluctuations:} Here, $\eta_\alpha < \eta_T$,
i.e., ${\cal D}_\alpha < 1$ and $|\eta_K|=+\eta_K=\eta_T-\eta_\alpha$, giving

\begin{eqnarray}
\frac{\gamma_>}{\sigma} &=& -\left(\frac{k}{k_\alpha}\right)^2+
\frac{{\cal D}_\alpha}{{1-\cal D}_\alpha}\left[
-\left(\frac{k}{k_\alpha}\right)^4+ {\cal D}_M
\left(\frac{k}{k_\alpha}\right)^2+
\frac{\vert\varepsilon_S\vert}{\sqrt{2}}\sqrt{{\cal R}_a(k)+{\cal R}_b(k)}
\right]
\label{gamweak}\\[2ex]
&&\mbox{with}\quad {\cal R}_a(k) = \left(\frac{k}{k_\alpha}\right)^2
\left[\left(\frac{k}{k_\alpha}\right)^2 - 1 \right]-
{\cal D}_M \left(\frac{k}{k_\alpha}\right)^2 \\[2ex]
&&\mbox{and}\quad {\cal R}_b(k) = \left(
\Bigg\{\left[\left(\frac{k}{k_\alpha}\right)^2 - 1 \right]
+ {\cal D}_M \left(\frac{k}{k_\alpha}\right)^2
\Bigg\}^2 + {\cal D}_M \left(\frac{k}{k_\alpha}\right)^2
\right)^{1/2}
\label{R}
\end{eqnarray}
\noindent
In the left panel of \Fig{fig2} we show the wavenumber dependence of
the normalized growth rate $\gamma_>/\sigma$ for different choices of
the shear parameter, at fixed ${\cal D}_\alpha$ and ${\cal D}_M$, when
$\alpha$ fluctuations are weak.
Interestingly, the growth rate is positive for fairly small
wavenumbers, thus facilitating a truly large scale dynamo, with a wavenumber
cutoff beyond which the growth rate turns negative. At much larger wavenumbers,
The growth rate varies as $\gamma \propto -k^4$ due to the
$\eta_T$--correction in the present model.
Shear boosts the growth
rates at all wavenumbers, and thus it can support the dynamo action for
sufficiently strong Moffatt drift.}

\paragraph{{\bf Strong $\alpha$ fluctuations:} In this case, $\eta_\alpha > \eta_T$,
i.e., ${\cal D}_\alpha > 1$ and $|\eta_K|=-\eta_K$, giving
\beq
\frac{\gamma_>}{\sigma} = +\left(\frac{k}{k_\alpha}\right)^2+
\frac{{\cal D}_\alpha}{{\cal D}_\alpha-1}\left[
-\left(\frac{k}{k_\alpha}\right)^4+ {\cal D}_M
\left(\frac{k}{k_\alpha}\right)^2+
\frac{\vert\varepsilon_S\vert}{\sqrt{2}}\sqrt{{\cal R}_a(k)+{\cal R}_b(k)}
\right]
\label{gamstr}
\eeq
Here the small wavenumbers grow as all the effects, Kraichnan
diffusivity, Moffatt drift and shear, contribute positively to the
dynamo action; see right panel of \Fig{fig2}. Similar to the case of weak $\alpha$ fluctuations, the growth rate here too is a non-monotonic function of $k$ and it becomes negative for sufficiently large wavenumbers.}

\subsubsection{Dynamo action for zero Moffatt drift}
This corresponds to the Kraichnan problem, extended to include a non-zero $\talp$. There are two cases to consider, the one in the absence of Shear and the other when Shear is present.

\medskip
\noindent
{\bf 1. Shear absent (only $\eta_\alpha$ and $\talp$ non zero)}

Using \Eqss{gamweak}{gamstr} by setting ${\cal D}_M=0$ and $|\varepsilon_S|=0$ the normalised growth rate can be expressed as, for

\paragraph{\bf Weak $\alpha$ fluctuations:} when $0<\eta_\alpha<\eta_T$, i.e., ${\cal D}_\alpha<1$,
\beq
\frac{\gamma}{\sigma} = -\left(\frac{k}{k_\alpha}\right)^2-\frac{{\cal D}_\alpha}{{1-\cal D}_\alpha}
\left(\frac{k}{k_\alpha}\right)^4 
\label{wkrfin}
\eeq 

Here, the growth is negative definite for all values of $k$ and monotonically decreasing function of $k$. At large wavenumbers, it varies as $\gamma\propto -k^4$, a correction due to finite $\talp$
and inclusion of finite resistive term in the fluctuating field equation.
The first term in the \Eq{wkrfin} is due to Kraichnan diffusivity (compare it with \Eq{greenfn} by setting $S=0$ and $\bfV_{\!\!M}=0$) 

\begin{figure}
\centering
\includegraphics[scale=0.4]{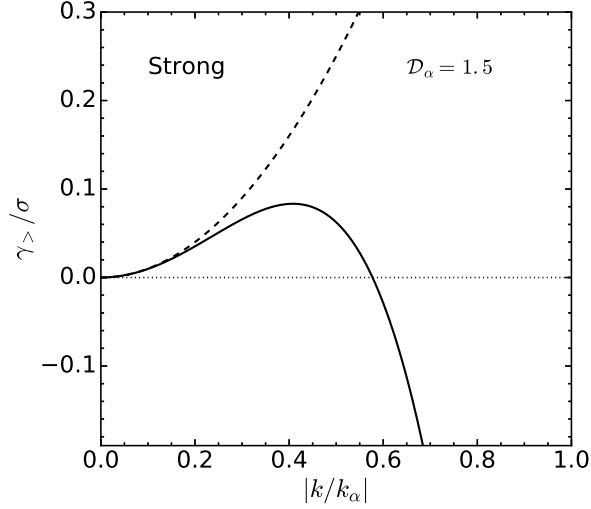}
\caption{Normalised growth rate when moffat drift and shear is zero. Plotted as a function of $|k/k_\alpha|$. Solid curve shows finite $\talp$ correction and dashed-dotted curve for white-noise case.}
\label{fig3}
\end{figure}

\paragraph{\bf Strong $\alpha$ fluctuations:} when $0<\eta_T<\eta_\alpha$, i.e., ${\cal D}_\alpha>1$,
\beq
\frac{\gamma}{\sigma} = \left(\frac{k}{k_\alpha}\right)^2-\frac{{\cal D}_\alpha}{{\cal D}_\alpha-1}
\left(\frac{k}{k_\alpha}\right)^4 
\eeq 
In this regime, the growth rate is positive for certain range of wavenumbers and it becomes negative for large wavenumbers as mentioned above. In \Fig{fig3} we compare our model (which has non--zero $\talp$)
with the original Kraichnan model --- we see that a non--zero $\talp$ introduces a high wavenumber cutoff in case of strong $\alpha$-fluctuations,
which agrees with the conclusions of \cite{S16}.

\medskip
\noindent
{\bf 2. The effect of Shear}

Using \Eq{Gamma} we rewrite the growth rate function more explicitly as:
\begin{eqnarray}
\Gamma_> \,&=&\, -\,\varepsilon_K\,-\,\beta^2\;, \qquad\qquad\qquad\qquad\qquad
\mbox{when}\quad0\,\leq\,\beta \,\leq\,1 
\label{Gamma-sd}\\[2ex]
\Gamma_> \,&=&\, -\,\varepsilon_K\,-\,\beta^2\,+\,
\vert\varepsilon_S\vert\;\sqrt{\beta(\beta - 1)}\;, \qquad
\mbox{when}\quad \beta \,>\,1.
\label{Gamma-sd2}
\end{eqnarray}
We note that the shear does not couple to the dynamo growth rate
when $\beta$ is smaller than or equal to unity, or in other words,
when $|k|<k_\alpha$\footnote{Compare the \Eqs{Gamma-sd}{Gamma-sd2} with equation (85) in SS14, where the authors had obtained the contribution of shear for $|k|<k_\alpha$, due to the error in the angle evaluation in \Eq{gamma} which is corrected here.}.
Since $\vert\varepsilon_S\vert \ll 1$, the dominant term in \Eq{Gamma-sd2}
is $-\beta^2$ for $k > k_\alpha$, it makes the growth rate as negative
definite. Thus, for weak $\alpha$ fluctuations which have
$\varepsilon_K>0$, shear alone cannot drive a large-scale
dynamo at any wavenumber. Therefore the necessary condition for
dynamo action in this case is that the $\alpha$ flcutuations must be strong.
We now look at the properties of growth rate as a function of wavenumber.

\paragraph{{\bf Weak $\alpha$ fluctuations:} Here, $\eta_\alpha < \eta_T$,
i.e., ${\cal D}_\alpha < 1$ and $|\eta_K|=+\eta_K=\eta_T-\eta_\alpha$, giving

\begin{eqnarray}
\frac{\gamma_>}{\sigma} &=& -\left(\frac{k}{k_\alpha}\right)^2-
\frac{{\cal D}_\alpha}{{1-\cal D}_\alpha}
\left(\frac{k}{k_\alpha}\right)^4,
\qquad\mbox{when}\quad 0<|k|<k_\alpha
\label{Gamw-sh} \\
 &=& -\left(\frac{k}{k_\alpha}\right)^2+
\frac{{\cal D}_\alpha}{{1-\cal D}_\alpha}\left[
-\left(\frac{k}{k_\alpha}\right)^4+\vert\varepsilon_S\vert\sqrt{\left(\frac{k}{k_\alpha}\right)^2\left[\left(\frac{k}{k_\alpha}\right)^2-1\right]}
\right],\nonumber\\
\mbox{when}\quad |k|>k_\alpha
\label{Gamw-sh2} 
\end{eqnarray}
We can see from \Eq{Gamw-sh} that the growth rate is negative definite in the range $0<|k|<k_\alpha$ as inferred above. Dynamo action is not 
possible for $|k|>k_\alpha$ for the following reason. When $|k|>k_\alpha$,  
shear contributes to the growth rate (see \Eq{Gamw-sh2}). Since the model is valid for $|\varepsilon_S|\ll 1$, in order to increase the strength of that term we can increase ${\cal D}_\alpha$ (while keeping it less than unity), but this will also strengthen the second term, which is $\propto -k^4$, due to finite $\eta_T$ correction in fluctuating field equation, thereby killing dynamo action.  } 
\begin{figure}
\centering
\includegraphics[width=0.49\columnwidth]{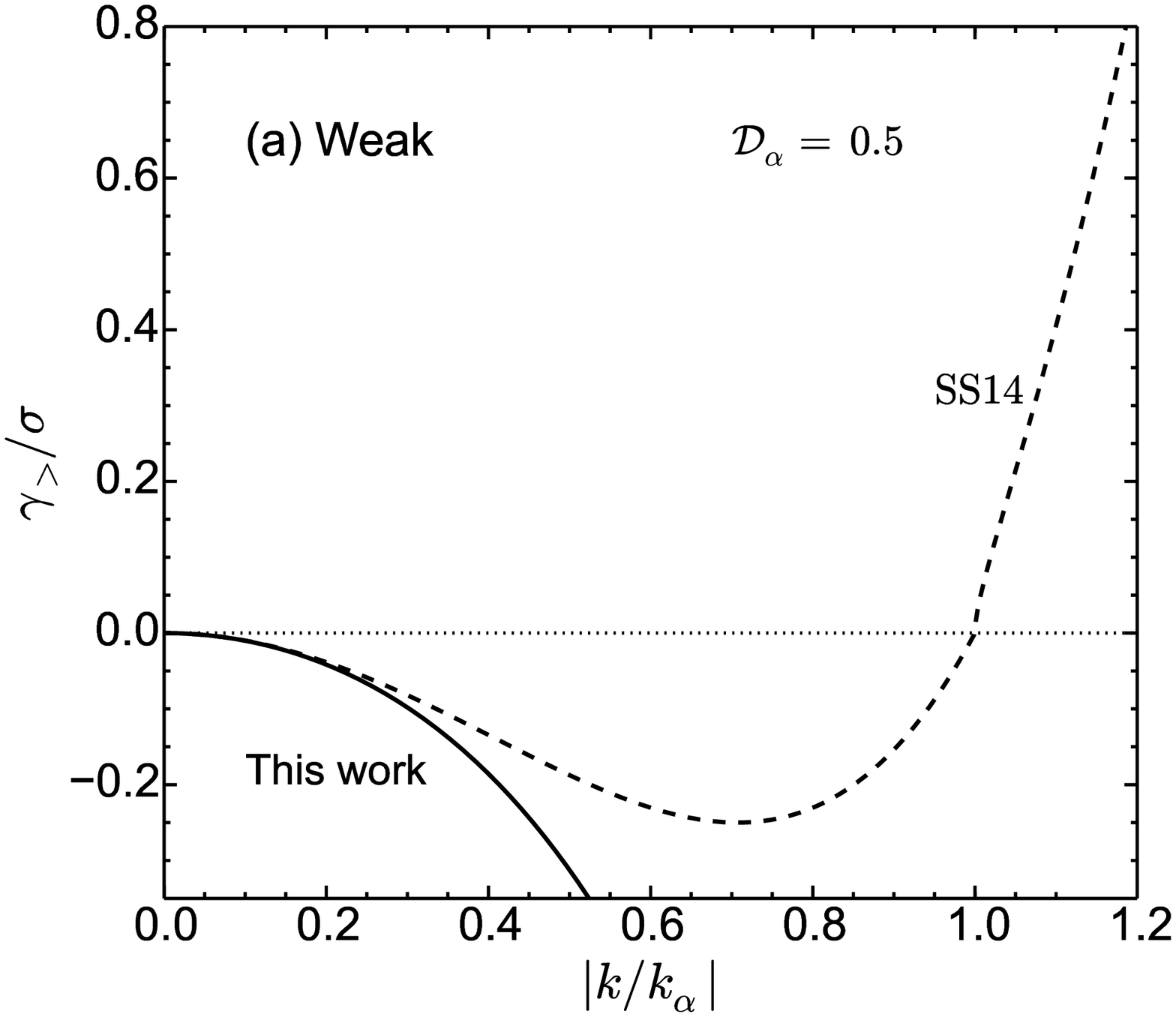}
\includegraphics[width=0.49\columnwidth]{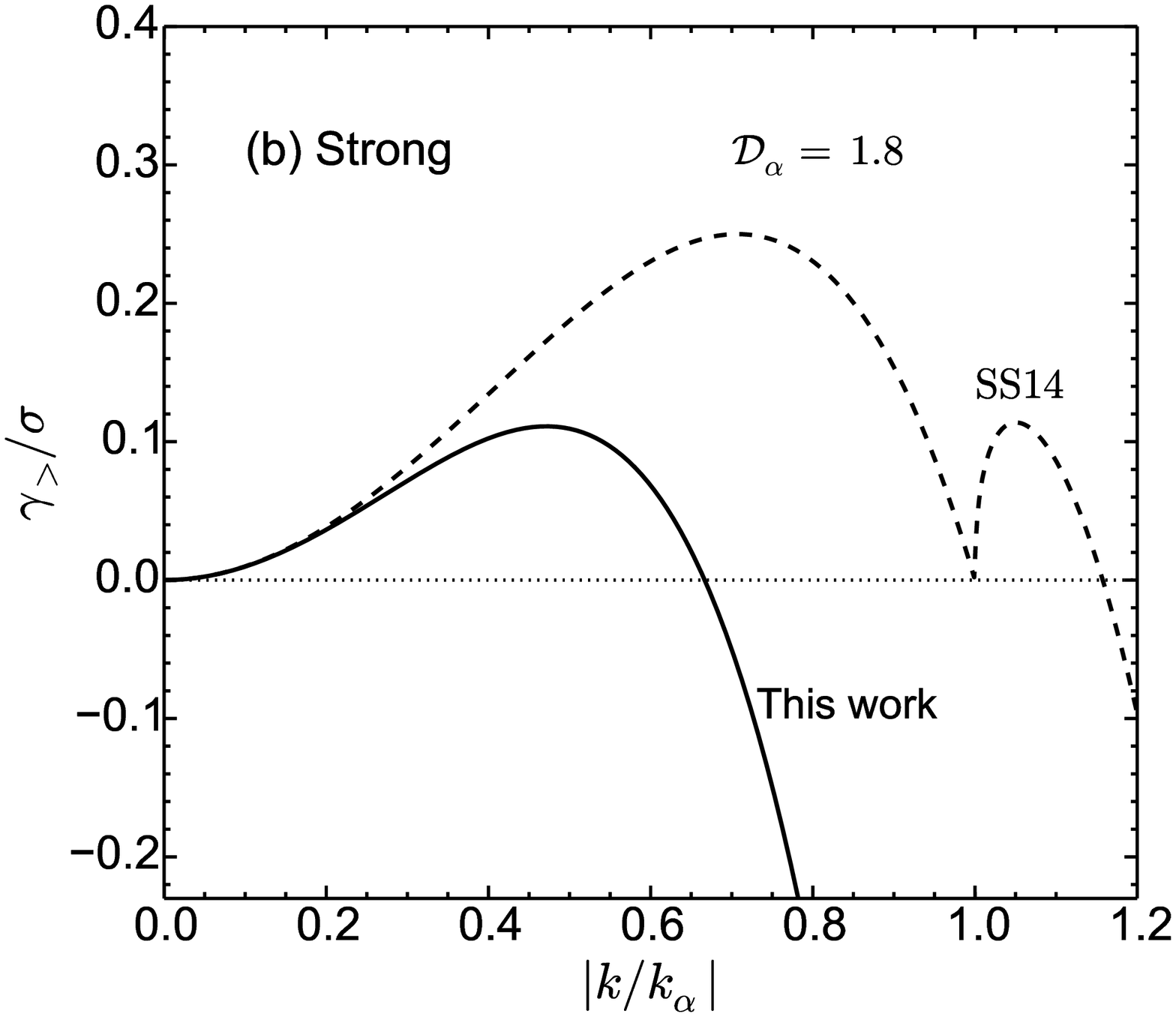}
\caption{
Normalized growth rate $\gamma_>/\sigma$ as function of $|k/k_\alpha|$
for $|\varepsilon_S|=0.3$. Left and right panels correspond to weak
(${\cal D}_\alpha=0.5$) and strong (${\cal D}_\alpha=1.8$) $\alpha$ fluctuations,
respectively. Solid and dashed curves correspond to this work and SS14, respectively.
}
\label{fig4}
\end{figure}

\noindent
{\bf Strong $\alpha$ fluctuations:} Here, $\eta_T < \eta_\alpha$,
i.e., ${\cal D}_\alpha > 1$ and $|\eta_K|=-\eta_K=\eta_\alpha-\eta_T$, giving

\begin{eqnarray}
\frac{\gamma_>}{\sigma} &=& \left(\frac{k}{k_\alpha}\right)^2-
\frac{{\cal D}_\alpha}{{\cal D}_\alpha-1}
\left(\frac{k}{k_\alpha}\right)^4,
\qquad\mbox{when}\quad 0<|k|<k_\alpha
\label{GamS-sh} \\
 &=& \left(\frac{k}{k_\alpha}\right)^2+
\frac{{\cal D}_\alpha}{{\cal D}_\alpha-1}\left[
-\left(\frac{k}{k_\alpha}\right)^4+\vert\varepsilon_S\vert\sqrt{\left(\frac{k}{k_\alpha}\right)^2\left[\left(\frac{k}{k_\alpha}\right)^2-1\right]}
\right],\nonumber \\
\mbox{when}\quad |k|>|k_\alpha|. 
\label{GamS-sh2}
\end{eqnarray}
The growth rate $\gamma$ becomes positive for the certain range of wavenumber depending upon the strength of $\alpha$-fluctuations, eventually becoming negative at large wavenumbers due to $-k^4$ term arising due to finite $\eta_T$ correction. This behaviour is compared in \Fig{fig4} with
\cite{SS14}; we can see that there is good agreement at low wavenumbers whereas at large wavenumbers there is a difference. The derivation of \cite{SS14} had neglected the effect of turbulent resistivity on the 
fluctuating component of the magnetic field, and they had noted that this
would lead to an overestimation of growth rates at large wavenumbers.
This is what we find in the present work: retaining this term makes 
growth rate negative at large wavenumbers and, for weak 
$\alpha$-fluctuations the behaviour is indeed qualitatively differently. Therefore, including $\eta_T$ term gives a bonafide large-scale dynamo action by predicting the high wavenumber cutoff. Thus, in the absence of Moffatt drift, the necessary condition for the large-scale dynamo when shear is present is same as the case when it is absent.

\section{Conclusions}

We have studied the effect of $\alpha$ fluctuations on the growth of 
large-scale magnetic fields in a shearing background. Our derivation of 
the mean electromotive force is based on the first-order smoothing approximation (FOSA), whose range of validity is given in \Eq{fosa-cond}.
These are such that FOSA is, in general, valid for all `weak' $\alpha$
fluctuations ($\eta_\alpha < \eta_T$), which is the case of primary 
interest for dynamo action. We have extended the analysis of \cite{SS14} by including the effect of the turbulent resistivity, $\eta_T$, on the  fluctuating component of the magnetic field. We derived the integro-differential equation for the large-scale magnetic field, which is non-perturbative in shear strength, $S$, and the $\alpha$-correlation time, $\talp$, similar to \cite{SS14}. For the exactly solvable case of white-noise $\alpha$-fluctuations dynamo action is possible only when $\alpha$-fluctuations are strong; this is also similar
to \cite{SS14}. In order to explore dynamo action in the regime of 
weak $\alpha$-fluctuations it is necessary to consider a non--zero $\talp$. Considering a small but non--zero $\talp$ and a slowly varying large--scale magnetic field, we reduced the integro-differential equation to a partial differential equation. We present an expression for mean EMF, correct upto first order in $\talp$. We also corrected an error in \cite{SS14} in the expression for the growth rate, $\gamma$. Our salient conclusions are listed below:
\medskip

\begin{itemize}
\item[a)] In the absence of Moffatt drift (i.e. $\bfV_{\!\!M}=0$) the 
growth rate is independent of shear when $0<|k|<k_\alpha$, and there is no dynamo action for weak $\alpha$-fluctuations even when $|k|>k_\alpha$ for
moderately small shear (i.e. $|\varepsilon_S|\ll 1$) --- see left panel of \Fig{fig4}. 
\medskip

\item[b)] For dynamo action with weak $\alpha$-fluctuations, it is
necessary that $\bfV_{\!\!M} \neq 0$: Moffatt drift couples strongly to shear and excites dynamo modes for $|k|<k_\alpha$ --- see item 6 in subsection~5.1 and \Fig{fig2}. 
\end{itemize}
\medskip

We briefly comment on different approaches adopted in some
related earlier works involving $\alpha$ fluctuations in a shearing
background. \citet{HMS11} considered tensorial
$\hat{\bf \alpha}$-fluctuations due to a quasi two dimensional
velocity field, whose dynamics is governed by the Navier-Stokes
equation at low Reynolds number, where the stochastic motions
occur due to a Gaussian random forcing which is delta-correlated
in time. A double averaging scheme was employed, first over
the `horizontal' (or $xy$) coordinates, and the second over the
statistics of the forcing function. They found that the first moment
of the magnetic field does not grow, while there is a growth of the
mean-squared magnetic field.
Note that the spatial fluctuations in $\hat{\bf \alpha}$
were ignored there, and the correlation time of only temporally fluctuating
$\hat{\bf \alpha}$ was assumed to be the same as that of the
velocity field.
\citet{MB12} also studied a model with tensorial $\hat{\bf \alpha}$
and allowed only temporal fluctuations which were further restricted
to be delta-correlated in time. When cross-correlations between
different $\hat{\bf \alpha}$ components were assumed to be zero,
they found growing solutions for the second moment of the mean magnetic
field, but not for the first moment.
However, when cross-correlations were allowed, large enough shear
promoted the growth of even the mean magnetic field.
Ignoring spatial structures and memory effects of the stochastic
$\alpha$ appear to be a serious limitation.
We remedy this in the present investigation where essential generalization
is made to explore new physical mechanisms driving the large--scale
dynamos, but by focussing here on the scalar $\alpha$ fluctuations to keep
the analysis simple. 

Thus, our model is a minimal extension of \cite{Kra76,moffatt1978},
where $\alpha$ is assumed to be a fluctuating pseudo-scalar field,
and $\eta_T$ is constant. We have constructed a model of large-scale
dynamo action with essential roles played by the Moffatt drift and
a non-zero correlation time. Hence our focus has been to keep the
tensorial structure of $\alpha$ as simple as possible, while exploring
the effect of spatio-temporal variations that are natural to turbulent
flows. We note that our work is almost completely complementary
to \citet{MB12}, wherein $\alpha$ fluctuations are tensorial but
have very restrictive space-time properties: no spatial variation
at all and with a zero correlation time. Indeed non-zero correlation
times and non-trivial spatial statistics appear essential for
dynamo action, as emphasised in item (b) above. We note here that
there seems to be some numerical evidence for pseudo-tensorial
$\alpha$ and tensorial $\eta_T$ fluctuations
\citep{BRRK08,SJ15,Rhein2014}. Our results, obtained for pseudo-scalar
$\alpha$, can be readily extended to tensorial fields.

Our analytical results for the growth rates of modes relies on a
perturbative expansion in $\talp$, which could also be generalised.
Another important assumption is the role of the shear in the
statistics of $\alpha$ fluctuations: these fluctuations have been
specified by a Galiean-invariant two-point correlation function
in factored form ${\cal A}(\bfR){\cal D}(t)$,
where $\bfR = \bfx-\bfx'+S t'(x_1-x_1')\ey$. Even though the
functional form of ${\cal A}$ has dependence on shear through
the argument $\bfR$, to the first order expansion in $\talp$,
neither ${\cal A}(0)=\eta_\alpha$ nor $\bfV_{\!\!M}$ depends
on shear explicitly. This is a limitation, since we can expect
a background shear flow to introduce anisotropy in the turbulent
flow which is the source of the fluctuations. Future modelling
must seek to be guided by numerical simulations that are designed
to measure the statistics of $\alpha$ fluctuations. 

%\begin{acknowledgements}
%Acknowledge ...
%\end{acknowledgements}

\appendix

\section{Derivation of \Eq{hsoln}}
\label{app1}

\noindent
Fourier transforming \Eq{fbfosa}, we obtain:
\begin{eqnarray}
\left(\frac{\partial}{\partial t}+\eta_T\,K^2\right)\widetilde{\bfh}
\;-\; S\widetilde{h}_1\ey
&\;=\;& \mathrm{i}\bfK\cross\widetilde{\bfM}\,,\qquad
\bfK\cendot\widetilde{\bfh} \;=\; 0\,,\qquad \widetilde{\bfh}(\bfk, 0) 
\;=\; \bf0\,,\nonumber\\[1em]
\mbox{with}\quad \bfK(\bfk, t) &\;=\;& 
\ex(k_1 - St\,k_2) \,+\, \ey k_2 \,+\, \ez k_3\,,\nonumber\\[1em]
\mbox{and}\quad \widetilde{\bfM}(\bfk, t) &\;=\;&  
\frac{1}{(2\pi)^3} \int \mathrm{d}^3k'\,{\widetilde{a}}^{\,*}(\bfk', t)\,
\widetilde{\overline{\bfH}}(\bfk+\bfk', t)\,.
\label{fbF}
\end{eqnarray}
We integrate \Eq{fbF} componentwise to write the following solution,
satisfying both constraints, $\bfK\cendot\widetilde{\bfh} 
= 0$ and $\widetilde{\bfh}(\bfk, 0) = \bf0\,$:
\begin{eqnarray}
\widetilde{\bfh}(\bfk, t) \;&=&\;  \int_0^t \mathrm{d}t'\,
\widetilde{G}_{\eta_T}(\bfk,t,t')
\left[\mathrm{i}\bfK(\bfk, t')\cross\widetilde{\bfM}(\bfk, t')\right]\;+
\nonumber \\
&&\;+\; \ey S\int_0^t \mathrm{d}t'\int_0^{t'} \mathrm{d}t''\,
\widetilde{G}_{\eta_T}(\bfk,t,t'')
\left[\mathrm{i}\bfK(\bfk, t'')\cross\widetilde{\bfM}(\bfk, t'')\right]_1\,.
\label{hsolna1}
\end{eqnarray}
The Green's function $\widetilde{G}_{\eta_T}(\bfk,t,t')$ is given in
\Eq{GF}, from where we can see a property that
$\widetilde{G}_{\eta_T}(\bfk,t,t')\times
\widetilde{G}_{\eta_T}(\bfk,t,t'')=\widetilde{G}_{\eta_T}(\bfk,t,t'')\,$,
which is used in getting \Eq{hsolna1}.
Reducing the double time integral in \Eq{hsolna1} to a single
time integral by using,
\beq
\int_0^t \mathrm{d}t'\int_0^{t'} \mathrm{d}t''\,f(t'') \;=\;
\int_0^t \mathrm{d}t' (t-t') f(t')\,,
\eeq
we obtain the FOSA solution for the fluctuating magnetic field as
given in \Eq{hsoln}.

\section{Two--point $\alpha$--correlator in Fourier space}
\label{app2}

\noindent
Here we derive a general expression for time--stationary
Galilean--invariant two--point $\alpha$--correlator
in Fourier space, where \Eq{aacorr} transforms to:
\begin{eqnarray}
&&\overline{\widetilde{a}(\bfk_1, t)\widetilde{a}^*(\bfk_3, t')} \;=\;
\int \mathrm{d}^3x_1 \mathrm{d}^3x_3 \exp{(-\mathrm{i}\,\bfk_1\cendot\bfx_1
+\mathrm{i}\,\bfk_3\cendot\bfx_3)}\;\overline{a(\bfx_1, t) a(\bfx_3, t')}\nonumber\\[2ex]
&&\;=\; 2{\cal D}(t-t')\int \mathrm{d}^3x_1 \mathrm{d}^3x_3 \exp{[-\mathrm{i}(\bfk_1\cendot\bfx_1
-\bfk_3\cendot\bfx_3)]}\,
{\cal A}\!\left(\bfx_1-\bfx_3+St'(x_{11}-x_{31})\ey\right)\,.
\nonumber
\end{eqnarray}
Using new integration variables, $\bfr=\bfx_1-\bfx_3\,$ and
$\bfr'=\frac{1}{2}(\bfx_1+\bfx_3)\,$, we get
\begin{eqnarray}
\overline{\widetilde{a}(\bfk_1, t)\widetilde{a}^*(\bfk_3, t')} &=&
2{\cal D}(t-t')\int \mathrm{d}^3r\, \mathrm{d}^3r' \exp{\left[\,-\mathrm{i}(\bfk_1-\bfk_3)
\cendot\bfr' - \frac{\mathrm{i}}{2}(\bfk_1+\bfk_3)\cendot\bfr\,\right]}
\; \times\nonumber\\[2ex]
&&\qquad\qquad\qquad\times\;{\cal A}\!\left(\bfr+St'r_1\ey\right)\nonumber\\[2ex]
&=& 2{\cal D}(t-t')(2\pi)^3\, \delta(\bfk_1-\bfk_3)\int \mathrm{d}^3r\,
\exp{(-\mathrm{i}\,\bfk_1\cendot\bfr)}\,{\cal A}\!\left(\bfr+St'r_1\ey\right)\,.
\nonumber
\end{eqnarray}
Another change of the integration variable to $\bfR=\bfr+St'r_1\ey\,$
gives us a compact form for the $2$--point correlator:
\begin{eqnarray}
\overline{\widetilde{a}(\bfk_1, t)\widetilde{a}^*(\bfk_3, t')} &\;=\;&
2{\cal D}(t-t')(2\pi)^3\, \delta(\bfk_1-\bfk_3)\,\widetilde{{\cal A}}\left(\bfK(\bfk_1, t')\right)\,,\nonumber\\[1em]
\mbox{where}\quad 
\widetilde{{\cal A}}(\bfK) &\;=\;& \int \mathrm{d}^3R\,
\exp{(-\mathrm{i}\,\bfK\cendot\bfR)}\; {\cal A}(\bfR)\,.
\label{2ptfou}
\end{eqnarray}
Note that $\widetilde{{\cal A}}(-\bfK) = \widetilde{{\cal A}}^*(\bfK)$
because ${\cal A}(\bfR)$ is a real function and that the argument of
the complex spatial power spectrum $\widetilde{{\cal A}}(\bfK)$ is
a time-dependent wavevector.

\section{Derivation of \Eqs{emfU}{Ufou}}
\label{app3}

\noindent
We derive an expression for the mean EMF in Fourier space by
using \Eq{hsoln} as:
\begin{eqnarray}
\widetilde{\overline{\bfE}}(\bfk, t) &=&   \int \mathrm{d}^3x
\exp{(-\mathrm{i}\,\bfk\cendot\bfx)}\,\overline{\bfE}(\bfx, t)
=\int \mathrm{d}^3x \exp{(-\mathrm{i}\,\bfk\cendot\bfx)}\,
\overline{a(\bfx, t)\,\bfh(\bfx, t)}\nonumber \\[2ex]
&=& \frac{1}{(2\pi)^3} \int \mathrm{d}^3k'\, \mathrm{d}^3k''\,
\delta(\bfk'+\bfk''-\bfk)\;
\overline{\widetilde{a}(\bfk', t)\,\widetilde{\bfh}(\bfk'', t)}
\nonumber \\[2ex]
&=& \frac{1}{(2\pi)^3} \int \mathrm{d}^3k'\, \mathrm{d}^3k''\,
\delta(\bfk'+\bfk''-\bfk)\;\int_0^t \mathrm{d}t'\,
\widetilde{G}_{\eta_T}(\bfk'',t,t')\;\times\nonumber\\[2ex]
&&\times\,\Biggl\{ 
\left[\mathrm{i}\bfK(\bfk'', t')\cross\overline{\widetilde{a}(\bfk', t)
\widetilde{\bfM}(\bfk'', t')}\right]
+\ey S(t-t')\left[\,\mathrm{i}\bfK(\bfk'', t')\cross
\overline{\widetilde{a}(\bfk', t)\widetilde{\bfM}(\bfk'', t')}\,\right]_1
\Biggr\}\nonumber\\
\label{emfF} 
\end{eqnarray}
This is given in terms of the quantity
$\overline{\widetilde{a}(\bfk', t)\widetilde{\bfM}(\bfk'', t')}$,
which can be determined by using the definition of $\widetilde{\bfM}$
from \Eq{fbF} and then using the time--stationary Galilean--invariant expression
for the two--point $\widetilde{a}$--correlator as given by \Eq{2ptfou}. We get
\begin{eqnarray}
\overline{\widetilde{a}(\bfk', t)\widetilde{\bfM}(\bfk'', t')} \;&=&\;
\frac{1}{(2\pi)^3} \int \mathrm{d}^3k'''\,
\overline{\widetilde{a}(\bfk', t)\widetilde{a}^*(\bfk''', t')}\;
\widetilde{\overline{\bfH}}(\bfk''+\bfk''', t')\nonumber\\[2ex]
&=&\;2\,{\cal D}(t-t')\,\widetilde{{\cal A}}\left(\bfK(\bfk', t')\right)\,
\widetilde{\overline{\bfH}}(\bfk'+\bfk'', t')\,.
\label{am}
\end{eqnarray}
Substituting \Eq{am} in \Eq{emfF} and solving the $k''$--integral using
the property of $\delta$--function, we immediately find the expression
for mean EMF as given in \Eq{emfU} in terms of a generalized complex velocity vector
$\widetilde{\bfU}$ defined by \Eq{Ufou}.

%\newpage

\bibliographystyle{jpp}

\bibliography{ms}

\begin{thebibliography}{42}
\expandafter\ifx\csname natexlab\endcsname\relax\def\natexlab#1{#1}\fi

\bibitem[Brandenburg {\em et~al.\/}(2008)Brandenburg, R{\"a}dler, Rheinhardt \&
  K{\"a}pyl{\"a}]{BRRK08}
{\sc Brandenburg, A., R{\"a}dler, K.-H., Rheinhardt, M. \& K{\"a}pyl{\"a},
  P.~J.} 2008 Magnetic diffusivity tensor and dynamo effects in rotating and
  shearing turbulence. {\em Astroph. J.\/} {\bf 676}~(1), 740.

\bibitem[Brandenburg \& Subramanian(2005)]{BS05}
{\sc Brandenburg, A. \& Subramanian, K.} 2005 Astrophysical magnetic fields and
  nonlinear dynamo theory. {\em Phys. Rep.\/} {\bf 417}~(1), 1--209.

\bibitem[{Han}(2017)]{JL17}
{\sc {Han}, J.~L.} 2017 {Observing Interstellar and Intergalactic Magnetic
  Fields}. {\em Annual Review of Astronomy and Astrophysics\/} {\bf 55},
  111--157.

\bibitem[Heinemann {\em et~al.\/}(2011)Heinemann, McWilliams \&
  Schekochihin]{HMS11}
{\sc Heinemann, T, McWilliams, JC \& Schekochihin, AA} 2011 Large-scale
  magnetic field generation by randomly forced shearing waves. {\em Physical
  review letters\/} {\bf 107}~(25), 255004.

\bibitem[Kleeorin \& Rogachevskii(2008)]{KR08}
{\sc Kleeorin, Nathan \& Rogachevskii, Igor} 2008 Mean-field dynamo in a
  turbulence with shear and kinetic helicity fluctuations. {\em Physical Review
  E\/} {\bf 77}~(3), 036307.

\bibitem[Kolekar {\em et~al.\/}(2012)Kolekar, Subramanian \& Sridhar]{KSS12}
{\sc Kolekar, Sanved, Subramanian, Kandaswamy \& Sridhar, S.} 2012 Mean-field
  dynamo action in renovating shearing flows. {\em Physical Review E\/} {\bf
  86}~(2), 026303.

\bibitem[Kolokolov {\em et~al.\/}(2011)Kolokolov, Lebedev \& Sizov]{Kol11}
{\sc Kolokolov, IV, Lebedev, VV \& Sizov, GA} 2011 Magnetic field correlations
  in random flow with strong steady shear. {\em Journal of Experimental and
  Theoretical Physics\/} {\bf 113}~(2), 339.

\bibitem[Kraichnan(1976)]{Kra76}
{\sc Kraichnan, Robert~H.} 1976 Diffusion of weak magnetic fields by isotropic
  turbulence. {\em Journal of Fluid Mechanics\/} {\bf 75}~(4), 657--676.

\bibitem[Krause \& R{\"a}dler(1980)]{krause1980}
{\sc Krause, F. \& R{\"a}dler, K.-H.} 1980 {\em Mean-Field Magnetohydrodynamics
  and Dynamo Theory\/}. Pergamon.

\bibitem[McWilliams(2012)]{McW12}
{\sc McWilliams, J.~C.} 2012 The elemental shear dynamo. {\em Journal of Fluid
  Mechanics\/} {\bf 699}, 414--452.

\bibitem[Mitra \& Brandenburg(2012)]{MB12}
{\sc Mitra, Dhrubaditya \& Brandenburg, Axel} 2012 Scaling and intermittency in
  incoherent $\alpha$--shear dynamo. {\em Monthly Notices of the Royal
  Astronomical Society\/} {\bf 420}~(3), 2170--2177.

\bibitem[Moffatt(1978)]{moffatt1978}
{\sc Moffatt, H.~K.} 1978 {\em Field Generation in Electrically Conducting
  Fluids\/}. Cambridge University Press.

\bibitem[Moffatt(1983)]{Mof83}
{\sc Moffatt, H.~K.} 1983 Transport effects associated with turbulence with
  particular attention to the influence of helicity. {\em Reports on Progress
  in Physics\/} {\bf 46}~(5), 621.

\bibitem[Nigro {\em et~al.\/}(2017)Nigro, Pongkitiwanichakul, Cattaneo \&
  Tobias]{Nig17}
{\sc Nigro, G, Pongkitiwanichakul, P, Cattaneo, F \& Tobias, SM} 2017 What is a
  large-scale dynamo? {\em Monthly Notices of the Royal Astronomical Society:
  Letters\/} {\bf 464}~(1), L119--L123.

\bibitem[Parker(1979)]{parker1979}
{\sc Parker, E.~N.} 1979 {\em Cosmical Magnetic Fields: Their Origin and Their
  Activity\/}. Oxford University Press.

\bibitem[Pongkitiwanichakul {\em et~al.\/}(2016)Pongkitiwanichakul, Nigro,
  Cattaneo \& Tobias]{Pon16}
{\sc Pongkitiwanichakul, Peera, Nigro, G, Cattaneo, F \& Tobias, SM} 2016
  Shear-driven dynamo waves in the fully nonlinear regime. {\em The
  Astrophysical Journal\/} {\bf 825}~(1), 23.

\bibitem[Proctor(2012)]{Pro12}
{\sc Proctor, MRE} 2012 Bounds for growth rates for dynamos with shear. {\em
  Journal of Fluid Mechanics\/} {\bf 697}, 504--510.

\bibitem[Proctor(2007)]{Pro07}
{\sc Proctor, Michael~RE} 2007 Effects of fluctuation on $\alpha$Ω dynamo
  models. {\em Monthly Notices of the Royal Astronomical Society: Letters\/}
  {\bf 382}~(1), L39--L42.

\bibitem[Rheinhardt {\em et~al.\/}(2014)Rheinhardt, Devlen, Rädler \&
  Brandenburg]{Rhein2014}
{\sc Rheinhardt, Matthias, Devlen, Ebru, Rädler, Karl-Heinz \& Brandenburg,
  Axel} 2014 Mean-field dynamo action from delayed transport. {\em Monthly
  Notices of the Royal Astronomical Society\/} {\bf 441}~(1), 116--126.

\bibitem[Richardson \& Proctor(2012)]{RP12}
{\sc Richardson, KJ \& Proctor, MRE} 2012 Fluctuating $\alpha$o dynamos by
  iterated matrices. {\em Monthly Notices of the Royal Astronomical Society:
  Letters\/} {\bf 422}~(1), L53--L56.

\bibitem[Rogachevskii \& Kleeorin(2007)]{rogachevskii2007}
{\sc Rogachevskii, I. \& Kleeorin, N.} 2007 Magnetic fluctuations and formation
  of large-scale inhomogeneous magnetic structures in a turbulent convection.
  {\em Phys. Rev. E\/} {\bf 76}~(5), 056307.

\bibitem[Rogachevskii \& Kleeorin(2008)]{RK08}
{\sc Rogachevskii, Igor \& Kleeorin, Nathan} 2008 Nonhelical mean-field dynamos
  in a sheared turbulence. {\em Astronomische Nachrichten\/} {\bf 329}~(7),
  732--736.

\bibitem[Ruzmaikin {\em et~al.\/}(1988)Ruzmaikin, Shukurov \&
  Sokoloff]{ruzmaikin1988}
{\sc Ruzmaikin, A., Shukurov, A. \& Sokoloff, D.} 1988 {\em Magnetic Fields of
  Galaxies\/}. Kluver Acad. Publ.

\bibitem[Silant'ev(2000)]{Sil00}
{\sc Silant'ev, N.~A.} 2000 Magnetic dynamo due to turbulent helicity
  fluctuations. {\em Astronomy and Astrophysics\/} {\bf 364}, 339--347.

\bibitem[Singh(2016)]{S16}
{\sc Singh, Nishant~K.} 2016 Moffatt-drift-driven large-scale dynamo due to
  $\alpha$ fluctuations with non-zero correlation times. {\em Journal of Fluid
  Mechanics\/} {\bf 798}, 696--716.

\bibitem[Singh \& Jingade(2015)]{SJ15}
{\sc Singh, Nishant~K. \& Jingade, Naveen} 2015 Numerical studies of dynamo
  action in a turbulent shear flow. i. {\em The Astrophysical Journal\/} {\bf
  806}~(1), 118.

\bibitem[Singh {\em et~al.\/}(2017)Singh, Rogachevskii \& Brandenburg]{SRB17}
{\sc Singh, Nishant~K, Rogachevskii, Igor \& Brandenburg, Axel} 2017
  Enhancement of small-scale turbulent dynamo by large-scale shear. {\em The
  Astrophysical Journal Letters\/} {\bf 850}~(1), L8.

\bibitem[Singh \& Sridhar(2011)]{SS11}
{\sc Singh, Nishant~K \& Sridhar, S} 2011 Transport coefficients for the shear
  dynamo problem at small reynolds numbers. {\em Physical Review E\/} {\bf
  83}~(5), 056309.

\bibitem[Sokolov(1997)]{Sok97}
{\sc Sokolov, D.~D.} 1997 The disk dynamo with fluctuating spirality. {\em
  Astronomy Reports\/} {\bf 41}~(1), 68--72.

\bibitem[Squire \& Bhattacharjee(2015a)]{SB15a}
{\sc Squire, J. \& Bhattacharjee, A.} 2015a Coherent nonhelical shear dynamos
  driven by magnetic fluctuations at low reynolds numbers. {\em The
  Astrophysical Journal\/} {\bf 813}~(1), 52.

\bibitem[Squire \& Bhattacharjee(2015b)]{SB15b}
{\sc Squire, Jonathan \& Bhattacharjee, Amitava} 2015b Generation of
  large-scale magnetic fields by small-scale dynamo in shear flows. {\em
  Physical review letters\/} {\bf 115}~(17), 175003.

\bibitem[Sridhar \& Singh(2010)]{SS10}
{\sc Sridhar, S. \& Singh, Nishant~K.} 2010 The shear dynamo problem for small
  magnetic reynolds numbers. {\em Journal of Fluid Mechanics\/} {\bf 664},
  265--285.

\bibitem[Sridhar \& Singh(2014)]{SS14}
{\sc Sridhar, S. \& Singh, Nishant~K.} 2014 Large-scale dynamo action due to
  $\alpha$ fluctuations in a linear shear flow. {\em Monthly Notices of the
  Royal Astronomical Society\/} {\bf 445}~(4), 3770--3787 (SS14).

\bibitem[Sridhar \& Subramanian(2009a)]{SS09a}
{\sc Sridhar, S. \& Subramanian, K.} 2009a Shear dynamo problem: Quasilinear
  kinematic theory. {\em Physical Review E\/} {\bf 79}~(4), 045305.

\bibitem[Sridhar \& Subramanian(2009b)]{SS09b}
{\sc Sridhar, S. \& Subramanian, K.} 2009b Nonperturbative quasilinear approach
  to the shear dynamo problem. {\em Physical Review E\/} {\bf 80}~(6), 066315.

\bibitem[Steenbeck {\em et~al.\/}(1966)Steenbeck, Krause \& R{\"a}dler]{SKR66}
{\sc Steenbeck, Max, Krause, F \& R{\"a}dler, K-H} 1966 Berechnung der
  mittleren lorentz-feldst{\"a}rke f{\"u}r ein elektrisch leitendes medium in
  turbulenter, durch coriolis-kr{\"a}fte beeinflu{\ss}ter bewegung. {\em
  Zeitschrift f{\"u}r Naturforschung A\/} {\bf 21}~(4), 369--376.

\bibitem[Sur \& Subramanian(2009)]{SurSub09}
{\sc Sur, Sharanya \& Subramanian, Kandaswamy} 2009 Galactic dynamo action in
  presence of stochastic $\alpha$ and shear. {\em Monthly Notices of the Royal
  Astronomical Society: Letters\/} {\bf 392}~(1), L6--L10.

\bibitem[Tobias \& Cattaneo(2013)]{TC13}
{\sc Tobias, Steven~M \& Cattaneo, Fausto} 2013 Shear-driven dynamo waves at
  high magnetic reynolds number. {\em Nature\/} {\bf 497}~(7450), 463.

\bibitem[Vishniac \& Brandenburg(1997)]{VB97}
{\sc Vishniac, Ethan~T \& Brandenburg, Axel} 1997 An incoherent
  $\alpha$-$\omega$ dynamo in accretion disks. {\em The Astrophysical
  Journal\/} {\bf 475}~(1), 263.

\bibitem[Yousef {\em et~al.\/}(2008b)Yousef, Heinemann, Rincon, Schekochihin,
  Kleeorin, Rogachevskii, Cowley \& McWilliams]{You08b}
{\sc Yousef, TA, Heinemann, T, Rincon, F, Schekochihin, AA, Kleeorin, N,
  Rogachevskii, I, Cowley, SC \& McWilliams, JC} 2008b Numerical experiments on
  dynamo action in sheared and rotating turbulence. {\em Astronomische
  Nachrichten\/} {\bf 329}~(7), 737--749.

\bibitem[Yousef {\em et~al.\/}(2008a)Yousef, Heinemann, Schekochihin, Kleeorin,
  Rogachevskii, Iskakov, Cowley \& McWilliams]{You08a}
{\sc Yousef, TA, Heinemann, T, Schekochihin, AA, Kleeorin, N, Rogachevskii, I,
  Iskakov, AB, Cowley, SC \& McWilliams, JC} 2008a Generation of magnetic field
  by combined action of turbulence and shear. {\em Physical review letters\/}
  {\bf 100}~(18), 184501.

\bibitem[Zeldovich {\em et~al.\/}(1983)Zeldovich, Ruzmaikin \&
  Sokolov]{zeldovich1983}
{\sc Zeldovich, Ya.~B., Ruzmaikin, A.~A. \& Sokolov, D.~D.} 1983 {\em Magnetic
  Fields in Astrophysics\/}. Gordon and Breach Science Publishers.

\end{thebibliography}

\end{document}